\documentclass[aps, pre, reprint, floatfix, longbibliography, superscriptaddress]{revtex4-1} 
\usepackage[utf8]{inputenc} 
\usepackage{amsfonts} 
\usepackage{amssymb} 
\usepackage{bm} 
\usepackage{color} 
\usepackage{dcolumn} 
\usepackage{float} 
\usepackage{mathrsfs} 
\usepackage{mathtools} 
\usepackage{scrextend} 
\usepackage[textsize=tiny]{todonotes} 
\newcommand{\p}{ \eta_{\mathrm{s}, y} }                        % the nondimensionalized pressure 
\newcommand{\pR}{ \eta_{\scriptscriptstyle \mathcal{R}} }      % the nondimensionalized pressure corresponding to the novel radius parameter $\mathcal{R}$ 
\newcommand{\Rm}{ R_{\scriptscriptstyle \mathrm{M}} }          % mean curvature radius 
\newcommand{\sigmam}{ \sigma_{\scriptscriptstyle \mathrm{M}} } % mean membrane stress 
\newcommand{\cp}{ \eta_\text{c} }                              % the nondimensionalized critical pressure 
\newcommand{\CP}{ p_\text{c} }                                 % the dimensionful critical pressure 
\newcommand{\CPv}{ p_{ \text{c}, \text{p} } }                  % the dimensionful critical pressure at the vertices of a spheroidal shell 
\newcommand{\ak}{ k_\text{asy} }                               % the asymptotic indentation stiffness 
\newcommand{\sk}{ k_\text{sph} }                               % stiffness of spherical shells 
\newcommand{\ck}{ k_\text{cyl} }                               % stiffness of cylindrical shells 
\newcommand{\ack}{ k_\text{asy, cyl} }                         % the asymptotic stiffness of cylindrical shells 
\newcommand{\ckzero}{ k_{\text{cyl}, 0} }                      % stiffness of cylindrical shells de Pablo 
\newcommand{\Iy}{ \frac{ { \ell_{\text{b}, y} }^2 }{\kappa} }  % the inverse of the stiffness scale corresponding to R_y 
\newcommand{\IR}{ \frac{ \mathcal{R} }{ \sqrt{ {\kappa}Y } } } % the inverse of the stiffness scale corresponding to the novel radius parameter 
\newcommand{\elly}{ \ell_{\text{b}, y} }                       % elastic scale using R_y 
\newcommand{\ellx}{ \ell_{\text{b}, x} }                       % elastic scale using R_x 
\newcommand{\ellp}{\ell_p}                                     % elastic scale using pressure 
\makeatletter
\DeclareFontFamily{U}{tipa}{}
\DeclareFontShape{U}{tipa}{m}{n}{<->tipa10}{}
\newcommand{\arc@char}{{\usefont{U}{tipa}{m}{n}\symbol{62}}}%

\newcommand{\arc}[1]{\mathpalette\arc@arc{#1}}

\newcommand{\arc@arc}[2]{%
  \sbox0{$\m@th#1#2$}%
  \vbox{
    \hbox{\resizebox{\wd0}{\height}{\arc@char}}
    \nointerlineskip
    \box0
  }%
}
\newcommand{\abs}[1]{ \left|{#1}\right| } % the absolute value of a number 
\newcommand{\imunit}{ \mathrm{i} } % √-1 
\newcommand{\Vlasov}{ \Delta_\text{V} }   % the Vlasov operator 
\newcommand{\FvK}{F\"{o}ppl-von K\'{a}rm\'{a}n } 
\newcommand{\secref}[1]{\textbf{\ref{#1}}} 
\newcommand{\eqnname}{Eq.} 
\newcommand{\eqnsname}{Eqs.}
\newcommand{\figref}[1]{\figurename{~\ref{#1}}}
\newcommand{\eqnref}[1]{\eqnname{~\eqref{#1}}}
\numberwithin{equation}{section} 
\renewcommand{\theequation}{\arabic{section}.\arabic{equation}} 
\begin{document} 
%%%%%%%%%%%%%%%%%%%%%%%%%%%%%%%%%%%%%%%%%%%%%%%%%%%%%%%%%%%%%%%%%%%%%%%%%%%%%%%%%%%%%%%%%%%%%%%%%%%%%%%%%%%%%%%%%%%%%%%%%%%%%%%%%%%%
\title{Indentation responses of pressurized ellipsoidal and cylindrical elastic shells: Insights from shallow-shell theory} 
% \title{Shallow-shell analysis of indentation responses of pressurized ellipsoidal and cylindrical elastic shells} 
\author{Wenqian Sun} 
\email[]{wenqians@uoregon.edu} 
\affiliation{Institute for Fundamental Science and Department of Physics, University of Oregon, Eugene, Oregon 97403, USA} 
\author{Jayson Paulose} 
\email[]{jpaulose@uoregon.edu} 
\affiliation{Institute for Fundamental Science and Department of Physics, University of Oregon, Eugene, Oregon 97403, USA} 
\affiliation{Material Science Institute, University of Oregon, Eugene, Oregon 97403, USA} 
%%%%%%%%%%%%%%%%%%%%%%%%%%%%%%%%%%%%%%%%%%%%%%%%%%%%%%%%%%%%%%%%%%%%%%%%%%%%%%%%%%%%%%%%%%%%%%%%%%%%%%%%%%%%%%%%%%%%%%%%%%%%%%%%%%%%
\date{\today} 
%%%%%%%%%%%%%%%%%%%%%%%%%%%%%%%%%%%%%%%%%%%%%%%%%%%%%%%%%%%%%%%%%%%%%%%%%%%%%%%%%%%%%%%%%%%%%%%%%%%%%%%%%%%%%%%%%%%%%%%%%%%%%%%%%%%%
\begin{abstract} 
Pressurized elastic shells are ubiquitous in nature and technology, from the outer walls of yeast and bacterial cells to artificial pressure vessels. 
Indentation measurements simultaneously probe the internal pressure and elastic properties of thin shells, and serve as a useful tool for strength testing and for inferring internal biological functions of living cells.
We study the effects of geometry and pressure-induced stress on the indentation stiffness of ellipsoidal and cylindrical elastic shells using shallow-shell theory. 
We show that the linear indentation response reduces to a single integral with two dimensionless parameters that encode the asphericity and internal pressure. 
This integral can be numerically evaluated in all regimes and is used to generate compact analytical expressions for the indentation stiffness in various regimes of technological and biological importance. 
Our results provide theoretical support for previous scaling and numerical results describing the stiffness of ellipsoids, reveal a new pressure scale that dictates the large-pressure response, and give new insights to the linear indentation response of pressurized cylinders. 
\end{abstract} 
%%%%%%%%%%%%%%%%%%%%%%%%%%%%%%%%%%%%%%%%%%%%%%%%%%%%%%%%%%%%%%%%%%%%%%%%%%%%%%%%%%%%%%%%%%%%%%%%%%%%%%%%%%%%%%%%%%%%%%%%%%%%%%%%%%%%
\maketitle 
%-----------------------------------------------------------------------------------------------------------------------------------
% \newpage 
\section{Introduction} \label{introduction} 
Thin curved shells are ubiquitous structures in nature and technology.
Their curvature inextricably links bending and stretching deformations, making them stiffer than flat plates of the same thickness and material---a phenomenon termed geometric rigidity~\cite{Audoly2010}.
The ability of closed shells to maintain a pressure difference between their interior and the environment also impacts their load-bearing properties, as is apparent from our everyday experience with balloons.
The interplay of elasticity, geometry, and pressure is crucial to our understanding of mechanical structures across a wide range of length scales, from viral capsids~\cite{Ivanovska2004} to reactor pressure vessels~\cite{Chattopadhyay2004}.

Indentation---gauging the deformation of a structure in response to a localized force---is a simple yet powerful tool for evaluating the mechanical properties of myriad structures~\cite{Cheng2004}, including shells.
Connecting shell indentation response to material properties and shape provides fundamental insight into geometric rigidity~\cite{Vaziri2008,Vella2012a,Lazarus2012a,Vella2012}, and is also of practical importance in evaluating the material properties of artificial~\cite{Gordon2004,Zoldesi2008} and biological~\cite{Arnoldi2000,Smith2000,Ivanovska2004,Deng2011} shell-like structures.
Although the general relation between indentation force and deflection is nonlinear and depends strongly on shell geometry, at small forces a linear regime can be identified in which the indentation force is  proportional to the inward displacement.
The constant of proportionality quantifies the \emph{indentation stiffness} of the shell, a metric which can be compared across geometries and size scales.
While theoretical analysis of the indentation stiffness of an unpressurized spherical shell dates back to the 1940s~\cite{Reissner1946}, few analytical results are available for other cases of interest.
For cylinders, the indentation stiffness is known in the unpressurized case~\cite{dePablo2003,Schaap2006}, and in the high-pressure limit ignoring bending rigidity~\cite{Arnoldi2000,Deng2011}.
An analytical expression for the indentation stiffness of internally-pressurized spherical shells was derived in Ref.~\onlinecite{Vella2012a}, and was subsequently generalized to external pressures~\cite{Paulose2012}.

For ellipsoids, a major advance was achieved in back-to-back experimental~\cite{Lazarus2012a} and theoretical~\cite{Vella2012} works reported in 2012.
Reference~\onlinecite{Lazarus2012a} proposed a form for the indentation stiffness of pressurized ellipsoids by analogy with known results for spheres, which was tested against experiments.
Reference~\onlinecite{Vella2012} used a perturbative analysis to obtain analytical results for the stiffness of nearly-spherical ellipsoidal shells, and combined this analysis with simulation results and physical scaling arguments to propose analytical forms for general ellipsoidal shells in the unpressurized and high-pressure limits. 
However, the relative contributions of the two geometric invariants describing a curved surface---the mean and Gaussian curvatures---was not rigorously established in these results. In addition, the focus on the tractable zero- and high-pressure limits leaves a gap in our theoretical understanding of the indentation stiffness of ellipsoids and cylinders at intermediate internal pressures. 

Here, we present a comprehensive theoretical analysis of the linear indentation stiffness of thin elastic ellipsoidal shells under both internal and external pressures.
Our main result is an expression for the indentation stiffness as an integral over a single variable, which includes the elastic moduli, curvature radii, and pressure as parameters.
The integral provides closed analytical forms in several limits, which agree with known results.
More generally, it can be numerically evaluated for arbitrary curvatures and pressures, providing a theoretical evaluation of the indentation stiffness in all regimes.
Conceptually, we provide a unifying framework which encompasses the local geometric rigidity of ellipsoids of arbitrary curvature including the spherical and cylindrical limits (up to important corrections at zero pressure in the cylinder limit, which we describe).
Besides providing analytical support to forms that were previously proposed using heuristic and scaling arguments~\cite{Vella2012,Lazarus2012a}, we also find a new pressure scale which controls the response of thin shells under high internal pressure, and obtain new expressions for the indentation stiffness of pressurized cylinders.

Our approach uses shallow-shell theory, which expresses the stress and displacement fields of a shallow section of the shell using Cartesian coordinates in a plane tangent to the indenting point.
Shallow-shell theory is widely used in elastic analyses of thin shells~\cite{Ventsel2002}, and provides an accurate description when the characteristic length scale of the deflection is small compared to the curvature radii.
As we will show, geometric rigidity ensures that point indentations induce such localized deflections at all pressures for non-cylindrical thin shells, and at non-zero internal pressures for cylindrical shells.
For the zero-pressure limit of cylindrical shells, the contribution of long-wavelength deflections becomes important, and the shallow-shell theory breaks down; different techniques are needed to understand the indentation stiffness of unpressurized cylinders, as has been done in Ref.~\onlinecite{dePablo2003}.
Nevertheless, we show that as internal pressure rises, shallow-shell theory becomes valid again, and provides useful new results above a threshold pressure which we derive. 
%-----------------------------------------------------------------------------------------------------------------------------------
% \newpage 
\section{Methods} \label{methods} 
In the current section, we will derive equations of equilibrium that characterize the local deformation of a given spheroidal shell, using the shallow-shell theory. 
We follow the presentation by Koiter and van der Heijden~\cite{Koiter_notes}. 
Further analysis of the equations of equilibrium will be included in \secref{results}. 

We start by mathematically describing an \textit{ellipsoidal} shell and the
deformation imposed on it, and then address the special case of spheroids which are ellipsoids of revolution.
The shells that we are interested in are thin, i.e., their thickness $t$ is much less than the other dimensions, so that they can be effectively treated as a two-dimensional surface. 
Besides encompassing a large class of artificial shells, thin-shell models have also been validated against experimental measurements for biological structures such as bacterial cell walls~\cite{Amir2014, Wong2017} and microtubules~\cite{Schaap2006}. 
%%%%%%%%%%%%%%%%%%%%%%%%%%%%%%%%%%%%%%%%%%%%%%%%%%%%%%%%%%%%%%%%%%%%%%%%%%%%%%%%%%%%%%%%%%%%%%%%%%%%%%%%%%%%%%%%%%%%%%%%%%%%%%%%%%%%
%%%%%%%%%%%%%%%%%%%%%%%%%%%%%%%%%%%%%%%%%%%%%%%%%%%%%%%%%%%%%%%%%%%%%%%%%%%%%%%%%%%%%%%%%%%%%%%%%%%%%%%%%%%%%%%%%%%%%%%%%%%%%%%%%%%%
\subsection{Description of Deformations of a Thin Shell} \label{framework} 
\begin{figure}[htb] 
\centering 
\includegraphics[width = 0.5 \textwidth]{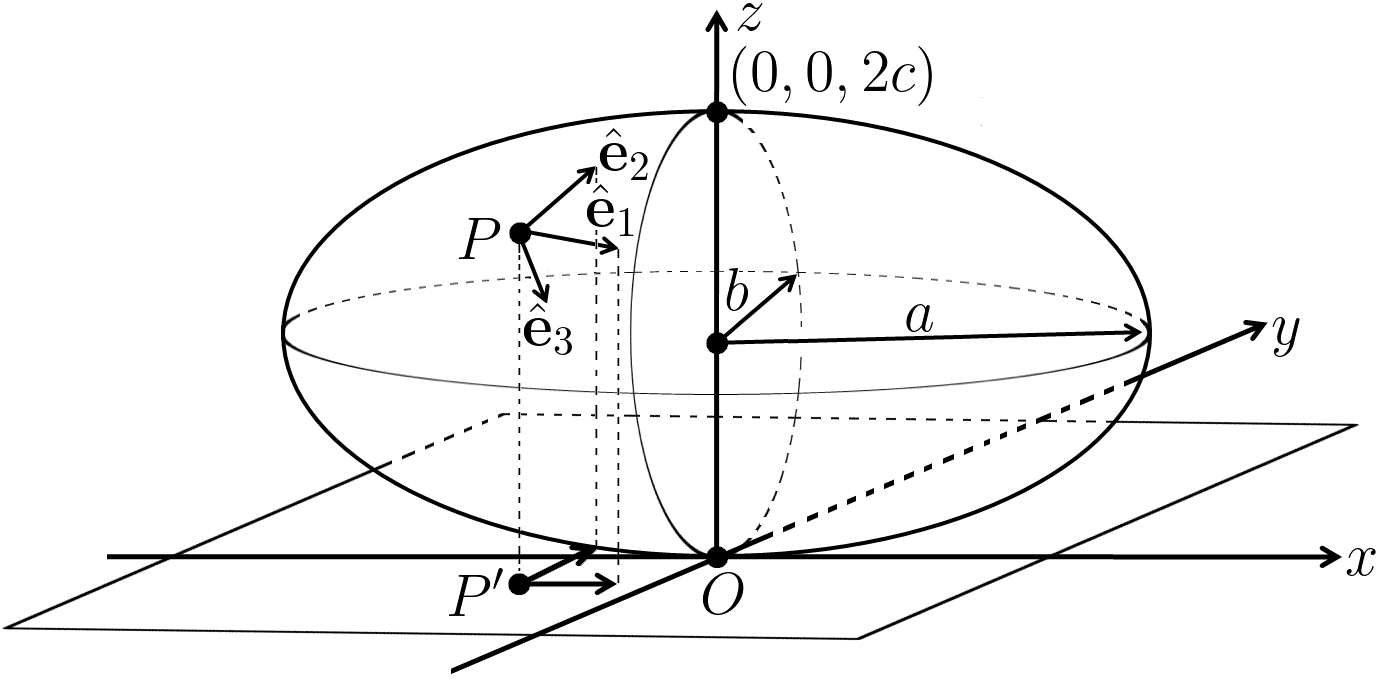} 
\caption{
A local coordinate representation of an ellipsoidal shell in the vicinity of the tangent point $O$. 
The two principal radii of curvature at $O$ are $R_x = \frac{a^2}{c}$ and $R_y = \frac{b^2}{c}$. 
For a spheroidal shell,  $b = c = R_y$ is the radius of the circular cross section at $O$.
Illustration depicts a prolate spheroid for which $R_x > R_y$; an oblate spheroid would correspond to $R_x < R_y$. 
The deformation at an arbitrary point $P$ is described by a displacement vector, which is decomposed into a non-orthonormal basis $\left\{\hat{ \mathbf{e} }_i\right\}_{i = 1}^3$. 
} 
\label{fig: an ellipsoid} 
\end{figure}

Let $O$ be one of the vertices of an ellipsoid where an external point load is exerted. (See \figurename{~\ref{fig: an ellipsoid}}.) 
We parametrize the ellipsoid using a right-hand Cartesian coordinate system whose $xy$-plane is the ellipsoid's tangent plane centered at $O$; the direction of the $x$-axis is chosen such that it coincides with the projection of the curve, corresponding to one of the two principal radii of curvature at $O$, onto the tangent plane. 
Accordingly, the $y$-axis will be automatically in-line with the projection associated with the other principal radius of curvature, and the $z$-axis points toward the center of the ellipsoid. 
\figurename{~\ref{fig: an ellipsoid}} illustrates such a coordinate system. 
A local coordinate representation of the ellipsoid can thus be written as 
$$
Z(x, y) = c - c\sqrt{1 - \left(\frac{x}{a}\right)^2 - \left(\frac{y}{b}\right)^2}. 
$$
For a shallow region of the ellipsoidal surface close to the origin such that 
$
\abs{ \frac{ \partial{Z} }{ \partial{x} }(x, y) }, \big|\frac{ \partial{Z} }{ \partial{y} }(x, y)\big| \ll 1, 
$
the expression above reduces to 
$$
Z \approx \frac{x^2}{2R_x} + \frac{y^2}{2R_y}, 
$$
where $R_x \coloneqq \frac{a^2}{c}$ and $R_y \coloneqq \frac{b^2}{c}$ are the two principal radii of curvature; it should be noted that in order for the shallow-shell assumption to hold, the approximated expression for $Z$ is only valid in a sufficiently small neighborhood of $O$. 

Points on the ellipsoid will get displaced under a deformation. 
The deformation can hence be described by a vector displacement field $\mathbf{u}(x, y)$ on the surface of the ellipsoid. 
Let $P = \big(x_0, y_0, Z(x_0, y_0)\big)$ be an arbitrary point on the surface. 
We decompose the displacement vector at $P$, $\mathbf{u}(x_0, y_0) \equiv \mathbf{u}_P$, as follows: 
\begin{align*} 
\mathbf{u}(x_0, y_0) &      = \sum_{i = 1}^3 u_i(x_0, y_0)\, \hat{ \mathbf{e} }_i(x_0, y_0) 
                              \\[0.25 em] 
                     & \equiv u_i(x_0, y_0)\, \hat{ \mathbf{e} }_i(x_0, y_0) 
                       \equiv u_{P, i}\, \hat{ \mathbf{e} }_{P, i}, 
\end{align*} 
where $\hat{ \mathbf{e} }_{P, 3}$ is the inward unit normal vector at $P$; $\hat{ \mathbf{e} }_{P, 1}$ and $\hat{ \mathbf{e} }_{P, 2}$ are unit vectors within the ellipsoid's tangent plane at $P$ chosen such that their projections onto the plane $Oxy$ coincide with $\hat{ \mathbf{x} }$ and $\hat{ \mathbf{y} }$, respectively. (See \figurename{~\ref{fig: an ellipsoid}}.) 
The Einstein summation convention was used. 
It should be pointed out that $\left\{\hat{ \mathbf{e} }_{P, i}\right\}_{i = 1}^3$ is generally not an orthonormal basis. 
The advantage of choosing such a basis is as follows: the prescribed deformation maps the point $P$ to 
$
\tilde{P} 
= 
\big( 
x_0 + \mathbf{u}_P \cdot \hat{ \mathbf{x} }, 
y_0 + \mathbf{u}_P \cdot \hat{ \mathbf{y} }, 
Z(x_0, y_0) + \mathbf{u}_P \cdot \hat{ \mathbf{z} } 
\big) 
\equiv 
\big( 
x + \mathbf{u} \cdot \hat{ \mathbf{x} }, 
y + \mathbf{u} \cdot \hat{ \mathbf{y} }, 
Z + \mathbf{u} \cdot \hat{ \mathbf{z} } 
\big)_P, 
$
which can be further approximated as 
$$
\left( 
x + u_1 - \frac{ \partial{Z} }{ \partial{x} } \cdot u_3, 
y + u_2 - \frac{ \partial{Z} }{ \partial{y} } \cdot u_3, 
Z + u_3 
\right)_P 
$$
under the assumptions that the shell is shallow (or, equivalently, $P$ is rather close to $O$), and $\abs{u_1}, \abs{u_2} \ll \abs{u_3}$ for all points on the shell. 

Once the deformation is mathematically characterized, the strain tensor $u_{\alpha\beta}(x, y)$ ($\alpha, \beta \in \{1, 2\}$) can be obtained by computing the change of metric, as~\cite{Koiter_notes} 
$$
u_{\alpha\beta} = \frac{1}{2}\left( 
                  \partial_\alpha{u_\beta} 
                + \partial_\beta{u_\alpha} 
                - \frac{2u_3}{ R_{\alpha\beta} } 
                + \partial_\alpha{u_3} \cdot \partial_\beta{u_3} 
                  \right), 
$$
where we have adopted the notations $\partial_1 \equiv \frac{\partial}{ \partial{x} }$ and $\partial_2 \equiv \frac{\partial}{ \partial{y} }$, and 
$$
\left[\frac{1}{ R_{\alpha\beta} }\right] 
\coloneqq 
\begin{pmatrix} 
\frac{1}{R_x} & 0 \\[0.25 em] 
0             & \frac{1}{R_y} 
\end{pmatrix} 
. 
$$
It should be pointed out that the derivation of the strain tensor assumes that the displacements are rapidly varying functions (on the scale of the shell's curvature radii) in the two principal directions, which is another key assumption of shallow-shell theory~\cite{Ventsel2002}. 
Mathematically, this means that $\abs{ \frac{1}{u_i}\partial_\alpha{u_i} } \gg \frac{1}{ \min\{R_x, R_y\} }$; for example, for wave-like deformations taking the form $u_i = u_{i, 0}e^{ \imunit\mathbf{q} \cdot \mathbf{r} }$, this criterion becomes $\frac{2\pi}{\lambda_\alpha} \eqqcolon q_\alpha \gg \frac{1}{ \min\{R_x, R_y\} }$, i.e., the deformation wavelength is much smaller than the principal radii of curvature. 
As we will see in \secref{the stiffness integral}, this assumption is well justified in the study of thin curved shells for a wide range of geometric parameters. 

For a two-dimensional isotropic elastic material, the stress tensor $\sigma_{\alpha\beta}(x, y)$ is related to the strain tensor via the strain-stress relation~\cite{LL_Elasticity} 
$$
\sigma_{\alpha\beta} = \frac{Et}{1 + \upsilon} 
                       \left(u_{\alpha\beta} + \frac{\upsilon}{1 - 2\upsilon}u_{\gamma\gamma}\delta_{\alpha\beta}\right), 
$$
where $E$ and $\upsilon$ denote the material's Young's modulus and Poisson's ratio, respectively; $\delta_{\alpha\beta}$ is the Kronecker delta, and, as aforementioned, $t$ stands for the thickness of the material; recall that for thin shells, $t \ll \min\{R_x, R_y\}$. 
These two tensor fields incorporate all the information about the deformation and will be used below to derive the elastic energy of the deformed shell. 
%%%%%%%%%%%%%%%%%%%%%%%%%%%%%%%%%%%%%%%%%%%%%%%%%%%%%%%%%%%%%%%%%%%%%%%%%%%%%%%%%%%%%%%%%%%%%%%%%%%%%%%%%%%%%%%%%%%%%%%%%%%%%%%%%%%%
%%%%%%%%%%%%%%%%%%%%%%%%%%%%%%%%%%%%%%%%%%%%%%%%%%%%%%%%%%%%%%%%%%%%%%%%%%%%%%%%%%%%%%%%%%%%%%%%%%%%%%%%%%%%%%%%%%%%%%%%%%%%%%%%%%%%
\subsection{Elastic Energy of the Deformed Shell} \label{energy functional} 
It is known from differential geometry that a closed shell cannot bend alone without stretching. 
If we further consider the closed shell to be pressurized, i.e., the shell is subjected to a pressure $p$, then the total elastic energy associated with the deformation will have three components, namely: bending energy, stretching energy and pressure energy (the work done by the pressure on the shell). 

In general, the stretching energy can be written as 
$$
E_\text{s}[ u_{\alpha\beta} ] = \frac{1}{2}\int_S \mathrm{d}{A}\, u_{\alpha\beta}\, \sigma_{\alpha\beta}. 
$$
The bending energy is related to the change of local curvatures and hence only depends on $u_3$~\cite{Ventsel2002}: 
\begin{align*} 
E_\text{b}[u_3] & = \frac{1}{2}\kappa\int_S \mathrm{d}{A}\, \Big\{\left(\Delta{u_3}\right)^2 
                    \\[0.25 em] 
                & \quad 
                  + 2(1 - \upsilon)\left[\left(\partial_{12}{u_3}\right)^2 - \partial_{11}{u_3} \cdot \partial_{22}{u_3}\right]\Big\}, 
\end{align*} 
where $\kappa = \frac{Et^3}{ 12(1 - \upsilon^2) }$ is the bending stiffness, and $\Delta \equiv \partial_{11} + \partial_{22}$ is the two-dimensional Laplacian operator. 
The pressure energy is simply given by 
$$
W[u_3] = -\int_S \mathrm{d}{A}\, pu_3; 
$$
the negative sign indicates our sign convention for the pressure on the shell that $p > 0$ ($p < 0$) corresponds to an internal (external) pressure. 
If an extra external point load with magnitude $F$ is acting at the point $O$, we can take into account the corresponding work done by simply replacing $p$ with $p' = p - F\delta^2( \mathbf{x} )$. 

The integration region $S$ is a portion of the ellipsoidal surface in the vicinity of $O$ outside of which the local deformation vanishes. 
Under the assumption that the shell is shallow, the area element $\mathrm{d}{A}$ gets simplified: 
$
\mathrm{d}{A}       = \sqrt{1 + \big(\frac{ \partial{Z} }{ \partial{x} }\big)^2 + \big(\frac{ \partial{Z} }{ \partial{y} }\big)^2} 
                      \, \mathrm{d}{x}\, \mathrm{d}{y} 
              \approx \mathrm{d}{x}\, \mathrm{d}{y}. 
$
Summing the bending, stretching and pressure energies, we obtain the total-energy functional 
\begin{align} 
\begin{split} 
E_\text{tot}[u_{\alpha\beta}, u_3] & \approx \int_{S'} \mathrm{d}{x}\, \mathrm{d}{y}\, 
                                             \Big\{\frac{1}{2}u_{\alpha\beta}\, \sigma_{\alpha\beta} + \frac{1}{2}\kappa\left(\Delta{u_3}\right)^2 
                                             \\[0.25 em] 
                                   & \ \ 
                                           + \kappa(1 - \upsilon)\left[\left(\partial_{12}{u_3}\right)^2 - \partial_{11}{u_3} \cdot \partial_{22}{u_3}\right] 
                                             \\[0.25 em] 
                                   & \quad 
                                           - p'u_3\Big\}, 
\end{split} 
\label{eqn: E_tot} 
\end{align} 
where $S'$ is the projected region of $S$ onto the plane $Oxy$. 
%%%%%%%%%%%%%%%%%%%%%%%%%%%%%%%%%%%%%%%%%%%%%%%%%%%%%%%%%%%%%%%%%%%%%%%%%%%%%%%%%%%%%%%%%%%%%%%%%%%%%%%%%%%%%%%%%%%%%%%%%%%%%%%%%%%%
%%%%%%%%%%%%%%%%%%%%%%%%%%%%%%%%%%%%%%%%%%%%%%%%%%%%%%%%%%%%%%%%%%%%%%%%%%%%%%%%%%%%%%%%%%%%%%%%%%%%%%%%%%%%%%%%%%%%%%%%%%%%%%%%%%%%
\subsection{Equations of Equilibrium (EOEs)} \label{EOEs} 
According to the variational principle, minimizing the total-energy functional gives a system of equations of equilibrium (EOEs) 
\begin{equation} 
\begin{cases} 
\kappa\Delta^2{u_3} + \Vlasov{\chi} - N_2(\chi, u_3) = p - F\delta^2( \mathbf{x} ) \\[0.25 em] 
\frac{1}{Y}\Delta^2{\chi} - \Vlasov{u_3} + \frac{1}{2}N_2(u_3, u_3) = 0, 
\end{cases} 
\label{eqn: non-linear EOEs} 
\end{equation} 
where the Vlasov operator in the shallow-shell theory is given by 
$
\Vlasov \equiv \frac{1}{R_y}\partial_{11} 
             + \frac{1}{R_x}\partial_{22}; 
$
$\chi(x, y)$ is the Airy stress function that encodes stress information in the following way: 
$$
\partial_{11}{\chi} = \sigma_{22}, \quad 
\partial_{22}{\chi} = \sigma_{11}, \quad \text{and} \quad 
\partial_{12}{\chi} = -\sigma_{12}; 
$$
$Y \coloneqq Et$, and $N_2(\cdot, \cdot)$ denotes a second-order nonlinear operator: for arbitrary twice-differentiable functions $f(x, y)$ and $g(x, y)$, 
$$
N_2\big(f, g\big) \coloneqq \partial_{11}{f} \cdot \partial_{22}{g} 
                          + \partial_{22}{f} \cdot \partial_{11}{g} 
                          - 2\partial_{12}{f} \cdot \partial_{12}{g}. 
$$

The next step is to linearize the nonlinear EOEs around the relaxed state of the shell in response to the uniform pressure.
Like the total-energy functional, a general deformation of pressurized elastic shells is a combination of two deformation states, a membrane state and a bending state. 
The membrane state describes the in-plane stresses that arise when the shell expands or contracts in response to the uniform internal or external pressure, with little change in local curvatures. 
The bending state describes the indentation responses due to the external point load, which lead to localized transverse deformations with a significant bending energy component.
To compute the linear indentation stiffness, we linearize the nonlinear EOEs around the membrane state, and solve the linearized equations for the bending state.

The membrane state of generalized ellipsoids is unwieldy and leads to distinct responses at all points on the ellipsoid. 
We therefore focus on the simpler case of spheroidal shells, which match most natural and artificial designs.
A spheroid is an ellipsoid of revolution, which has circular cross sections along one of its principal axes. 
We take $Ox$ to be that axis, and $b=c=R_y$ is thus the radius of the circular cross section at $O$, which corresponds to the equator of the spheroid. (See \figurename{~\ref{fig: an ellipsoid}}.)
The indentation stiffness we compute applies to all points on the equator, which are geometrically identical.

The Airy stress function corresponding to the membrane state for points on a spheroid's equator (the boundary of the spheroid's largest circular cross section) is known as~\cite{Clark1957} 
\begin{equation}
  \label{eq:airyfn}
\chi_0(x, y) = \frac{1}{4}pR_y\left[y^2 + \left(2 - \frac{R_y}{R_x}\right)x^2\right].   
\end{equation}

We can thus write an ansatz that corresponds to spheroidal shells as 
$$
\begin{cases} 
u_3(x, y) = u_{3, 0} + u_{3, 1}(x, y) \\[0.25 em] 
\chi(x, y) = \chi_0 + \chi_1(x, y), 
\end{cases} 
$$
where the subscripts $0$ and $1$ denote the membrane and the bending states, respectively. 
Note that we took the normal displacement field in the membrane state, $u_{3, 0}$, as a constant, which is only valid for points on the spheroid's equator. 
Substituting this ansatz into \eqnname{~\eqref{eqn: non-linear EOEs}} and discarding the terms quadratic in $u_{3, 1}$ and $\chi_1$ that are assumed to be small, we finally obtain the linearized EOEs for spheroidal shells, 
\begin{widetext} 
\begin{equation} 
\begin{cases} 
\kappa\Delta^2{ u_{3, 1} } + \Vlasov{\chi_1} 
- \frac{1}{2}pR_y\Delta{ u_{3, 1} } - \frac{1}{2}pR_y\left(1 - \frac{R_y}{R_x}\right)\partial_{22}{ u_{3, 1} } 
= -F\delta^2( \mathbf{x} ) \\[0.25 em] 
\frac{1}{Y}\Delta^2{\chi_1} - \Vlasov{ u_{3, 1} } = 0. 
\end{cases} 
\label{eqn: linear EOEs} 
\end{equation} 
\end{widetext} 
Although we have related the elastic moduli $\kappa$ and $Y$ to microscopic constants of uniform materials, they may also be regarded as effective moduli that penalize changes in metric and curvature of more complex quasi-two-dimensional structures such as biological shells. 

The physical quantity that we seek by solving this set of differential equations is the indentation stiffness. 
Recall that a point load with the magnitude $F$ is acting at $O$, an arbitrary point on a spheroid's equator that serves as the origin in our setup, and $u_{3, 1}(x, y)$ describes the out-of-plane deformation due to the point load. 
The indentation stiffness is then dictated by the Hooke's law: 
\begin{equation*} 
k = -\frac{F}{ u_{3, 1}(\mathbf{x} = 0) }, 
\end{equation*} 
where the fact that the indentation stiffness is a local property has allowed us to set $\mathbf{x} = 0$. 
%-----------------------------------------------------------------------------------------------------------------------------------
% \newpage 
\section{Results and Discussion} \label{results} 
%%%%%%%%%%%%%%%%%%%%%%%%%%%%%%%%%%%%%%%%%%%%%%%%%%%%%%%%%%%%%%%%%%%%%%%%%%%%%%%%%%%%%%%%%%%%%%%%%%%%%%%%%%%%%%%%%%%%%%%%%%%%%%%%%%%%
%%%%%%%%%%%%%%%%%%%%%%%%%%%%%%%%%%%%%%%%%%%%%%%%%%%%%%%%%%%%%%%%%%%%%%%%%%%%%%%%%%%%%%%%%%%%%%%%%%%%%%%%%%%%%%%%%%%%%%%%%%%%%%%%%%%%
\subsection{The Stiffness Integral} \label{the stiffness integral} 
We have derived the EOEs that characterize the local deformation around $O$, \eqnname{~\eqref{eqn: linear EOEs}}, which is a system of coupled linear partial differential equations (PDEs). 
Fourier transform can be applied to solve linear PDEs; the resulting solutions always take the form of an integral. 
The integral often cannot be solved analytically. 
However, in our case, the local nature of the probe and the measurement leads to a simplified Fourier integral.
Recall that the physical quantity that we are after is the local indentation stiffness which only requires the knowledge of the normal displacement at the origin due to the point load, $u_{3, 1}(\mathbf{x} = 0)$. 
This local nature avoids the need to solve \eqnname{~\eqref{eqn: linear EOEs}} over the entire domain, and enables analytical evaluation of the integral solution in our system, at least in certain limits. 

Our convention for Fourier transform is the following: for any Fourier-transformable function of two variables $f(x, y) \equiv f( \mathbf{x} )$, 
$$
f( \mathbf{x} ) = \int \! \frac{ \mathrm{d}^2{q} }{ (2\pi)^2 }\, \hat{f}( \mathbf{q} )e^{ \imunit\mathbf{q} \cdot \mathbf{x} }, 
$$
where 
$$
\hat{f}( \mathbf{q} ) \equiv \mathscr{F}\{f( \mathbf{x} )\}( \mathbf{q} ) 
                           = \int \! \mathrm{d}^2{x}\, f( \mathbf{x} )e^{ -\imunit\mathbf{q} \cdot \mathbf{x} }. 
$$
Performing Fourier transform of \eqnname{~\eqref{eqn: linear EOEs}} and combining the resulting two algebraic equations gives 
$$
F = \mathscr{F}\{F\delta^2( \mathbf{x} )\}( \mathbf{q} ) 
  = -\tilde{k}( \mathbf{q} ) \cdot \hat{u}_{3, 1}( \mathbf{q} ), 
$$
where 
\begin{align} 
\begin{split} 
\tilde{k}(\mathbf{q}) & \coloneqq {\kappa}q^4 + \frac{Y}{q^4}\left(\frac{ {q_x}^2 }{R_y} + \frac{ {q_y}^2 }{R_x}\right)^2 
                                  \\[0.25 em] 
                      & \ \ 
                                + \frac{1}{2}pR_yq^2 + \frac{1}{2}pR_y\left(1 - \frac{R_y}{R_x}\right){q_y}^2. 
\end{split} 
\label{eqn: Fourier stiffness} 
\end{align} 
The indentation stiffness can thus be written as 
\begin{equation} 
k = -\frac{F}{ u_{3, 1}(\mathbf{x} = 0) } 
  = \left[\int \! \frac{ \mathrm{d}^2{q} }{ (2\pi)^2 }\, \frac{1}{ \tilde{k} }\right]^{-1} 
  \eqqcolon 
    I^{-1}. 
\label{eqn:stiffness_integral} 
\end{equation} 

\eqnsname{~\eqref{eqn: Fourier stiffness}} and \eqref{eqn:stiffness_integral} underpin all our forthcoming results: the problem of computing the indentation stiffness has been reduced to evaluating a real-valued integral $I$, which we term the \emph{stiffness integral}, over the two-dimensional wavevector space.
However, we have yet to specify the integration limits in \eqnref{eqn:stiffness_integral}.
As with any physical theory, the integration must strictly be carried out over some range of wavevectors $\mathbf{q}$ for which the Fourier-transformed stiffness (\eqnref{eqn: Fourier stiffness}) is valid.
The large-wavevector, or UV, cutoff is dictated by the smallest wavelength for which shallow-shell theory is valid, which is of order the shell thickness $t$.
We will see that the integrand falls to zero for wavevectors much smaller than $1/t$, so the upper limit of the integration can be safely taken to $q = |\mathbf{q}| \to \infty$ (i.e., the theory is UV-convergent).

The treatment of the small-wavelength, or IR, cutoff requires more care.
The deflection-strain relations used in shallow-shell theory are accurate only for deflections which vary over length scales that are small compared to the radii of curvature, i.e., \eqnref{eqn: Fourier stiffness} is strictly valid only for $\abs{q_x} \gtrsim 1/R_x$ and $\abs{q_y} \gtrsim 1/R_y$. 
Nonetheless, the physics of deflection of thin curved shells allows us to take the small-wavevector limit to $q = 0$ in the stiffness integral without sacrificing accuracy for a wide range of geometries, provided the shells are thin.
To understand why, consider the Fourier contributions to the stiffness integral when $p=0$.
From \eqnref{eqn: Fourier stiffness}, we find that $1/\tilde{k}(\mathbf{q})$ has a roughly even contribution over a region in Fourier space within the bounds $|q_x| \lesssim 1/\elly$ and $|q_y| \lesssim 1/\ellx$, where $\ellx= \sqrt[4]{ \frac{ {\kappa}{R_x}^2 }{Y} }$ and $\elly= \sqrt[4]{ \frac{ {\kappa}{R_y}^2 }{Y} }$ are two elastic length scales arising from the balance between bending and stretching.
For thin shells, $\ellx \sim \sqrt{R_x t}$ and $\elly \sim \sqrt{R_y t}$ scale with the geometric mean of the curvature radii and the thickness, and are therefore small compared to the curvature radii themselves yet large compared to the shell thickness which serves as the UV cutoff scale for the theory.
This separation of length scales---a consequence of the interplay of geometry and elasticity---is responsible for the success of shallow-shell theory for understanding the indentation of thin shells, as has previously been recognized for spherical shells~\cite{Koiter1963,Paulose2012}.

For a broad range of thin-shell spheroidal geometries satisfying $ \sqrt{R_y t} \ll R_x \ll R_y^2/t$, the stiffness integral at zero pressure is dominated by modes with wavevectors in the range $1/R_x \lesssim |q_x| \lesssim 1/\elly$ and $1/R_y \lesssim |q_y| \lesssim 1/\ellx$.
As a result, including the erroneous but finite contributions to the integral for wavevectors near the origin ($|q_x| \lesssim 1/R_x$, $|q_y| \lesssim 1/R_y$) introduces an insignificant error to the indentation stiffness and the lower limit of integration can be taken to $q \to 0$ for these shells. 
However, the required separation of scales breaks down when $R_x \to 0$ (extremely narrow oblate shells) or $R_x \to \infty$ (cylinders) and the stiffness integral becomes invalid at zero pressure in these limits.
At finite internal and external pressures, the convergence of the stiffness integral depends on additional physical considerations.
We will address these considerations separately in the remainder of this subsection (where we impose the more stringent lower limit  $R_x \geq R_y/2$ on the curvature along the $x$ direction), as well as in sections \secref{buckling} (which discusses the behavior of the stiffness integral under external pressure) and \secref{cylindrical shells} (which revisits the integral for internally-pressurized cylinders).
In the latter subsection, we show that the indentation of cylindrical shells with a finite internal pressure is successfully captured by shallow-shell theory even though the criterion $R_x \ll R_y^2/t$ is violated.
Through these investigations, we will identify ranges of pressure values for which the stiffness integral, \eqnref{eqn:stiffness_integral} with lower and upper limits $q=0$ and $q=\infty$ respectively, accurately captures the indentation stiffness for \emph{all} spheroidal shell geometries with $R_x \geq R_y/2$ upto and including the cylinder limit of $R_x \to \infty$. 

Next, we nondimensionalize the stiffness integral using appropriate physical scales.
Of the two elastic length scales in the problem, we choose the scale $\ell_{\text{b}, y} = \sqrt[4]{ \frac{ {\kappa}{R_y}^2 }{Y} }$ associated with the equatorial radius to rescale lengths.
For the pressure scale, we choose $p_\text{sc} = \frac{ 4\sqrt{ {\kappa}Y } }{ {R_y}^2 }$, the absolute value of the buckling pressure of a spherical shell with the same equatorial radius~\cite{zoelly1915ueber}. 
Also, observing that parts of $\tilde{k}$ depend on $q = \sqrt{ {q_x}^2 + {q_y}^2 }$, we rewrite the stiffness integral in polar coordinates. 
Accordingly, after some algebra, we obtain 
\begin{widetext} 
\begin{equation} 
I(\beta, \p) = \frac{1}{8\pi^2}\Iy\int_0^{2\pi} \mathrm{d}{\theta}\, 
               \int_0^{+\infty} 
               \frac{ \mathrm{d}{x} } 
               { 
               \left[x + \p\left(1 + \beta\sin^2{\theta}\right)\right]^2 
             + \left[\left(1 - \beta\sin^2{\theta}\right)^2 - \p^2\left(1 + \beta\sin^2{\theta}\right)^2\right] 
               }. 
\label{eqn: 2D stiffness integral} 
\end{equation} 
\end{widetext} 
Thereinto, $\p \coloneqq \frac{p}{ p_\text{sc} } = \frac{p{R_y}^2}{ 4\sqrt{ {\kappa}Y } }$ is the scaled pressure, and our sign convention for the background pressure $p$ carries over: a positive (negative) $\p$ corresponds to an internal (external) pressure. 
The geometry of the spheroid is captured in the parameter $\beta \coloneqq 1 - R_y / R_x$ which characterizes the asphericity of a given spheroidal shell; specifically, spheroids with $\beta > 0$  are prolate whereas $\beta < 0$ corresponds to oblate spheroids. (See \figurename{~\ref{fig: an ellipsoid}}.)
Moreover, for a prolate spheroid, $\sqrt{\beta} = \sqrt{ 1 - \frac{b^2}{a^2} } = \varepsilon$ is in fact the eccentricity of its elliptical cross sections. 

A few geometries  are of special interest.
For spherical shells, $\beta = \varepsilon = 0$ ($R_x = R_y$), i.e., both cross sections at $O$ are circular. 
At the other extreme, infinitely long, circular cylindrical shells have $\beta = \varepsilon = 1$ ($R_x \to \infty$), i.e., the elliptical cross section at $O$ becomes an unbound rectangle with width $R_y$. 
The oblate spheroid $\beta = -1$ does not appear to have a special geometry, but is important for stability reasons due to the form of the membrane prestresses.
When $\beta < -1$, $R_y < 2R_x$, and according to \eqnref{eq:airyfn}, the in-plane stress along $y$-direction in the membrane state, $\sigma_{22}^0 = \partial_{11}{\chi_0}$ is \emph{negative}, i.e., compressive, for \emph{internally} pressurized shells.
As a result, oblate shells with $\beta \leq -1$ may even buckle under an internal pressure~\cite{Internal_Buckling}.
Correspondingly, the positivity of the Fourier-transformed stiffness, \eqnref{eqn: Fourier stiffness}, cannot be guaranteed at all wavevectors even for positive pressures, and the corresponding analysis of the stiffness integral will be hence somewhat different.
In the rest of the paper, we will restrict our focus to shells with $-1 < \beta \leq 1$. 

The indentation stiffness due to some general load function (that specifies the spatial distribution of external forces), instead of a point load, can be calculated as the spatial convolution of $k$ and the load function since $k$ can be thought of as a Green's function. 
Such a convolution integral might be carried out numerically, as long as the load function is itself confined to the shallow region of interest. 
% The domain of the load function must be a subset of $S'$. 
%%%%%%%%%%%%%%%%%%%%%%%%%%%%%%%%%%%%%%%%%%%%%%%%%%%%%%%%%%%%%%%%%%%%%%%%%%%%%%%%%%%%%%%%%%%%%%%%%%%%%%%%%%%%%%%%%%%%%%%%%%%%%%%%%%%%
%%%%%%%%%%%%%%%%%%%%%%%%%%%%%%%%%%%%%%%%%%%%%%%%%%%%%%%%%%%%%%%%%%%%%%%%%%%%%%%%%%%%%%%%%%%%%%%%%%%%%%%%%%%%%%%%%%%%%%%%%%%%%%%%%%%%
\subsection{Zero-Pressure Stiffness} \label{zero pressure} 
As a direct check, we can first calculate the indentation stiffness at zero pressure with the stiffness integral. 
Setting $\p = 0$ reduces \eqnname{~\eqref{eqn: 2D stiffness integral}} to 
\begin{align*} 
I(\beta, \p = 0) & = \frac{1}{8\pi^2}\Iy\int_0^{2\pi} \mathrm{d}{\theta}\, 
                     \\[0.25 em] 
                 & \ \ \times 
                     \int_0^{+\infty} \frac{ \mathrm{d}{x} }{x^2 + \left(1 - \beta\sin^2{\theta}\right)^2} 
                     \\[0.25 em] 
                 & = \frac{1}{8}\Iy\frac{1}{ \sqrt{1 - \beta} } 
                   = \frac{1}{8}\sqrt{ \frac{R_xR_y}{ {\kappa}Y } }. 
\end{align*} 
Taking the inverse then gives the zero-pressure stiffness 
\begin{equation} 
k_{p = 0} = 8\sqrt{ {\kappa}Y }\sqrt{K}. 
\label{eqn: zero pressure stiffness} 
\end{equation} 
This agrees with the conjecture in Ref.~\onlinecite{Vella2012}, analytically showing that the zero-pressure stiffness is indeed governed by a shell's local Gaussian curvature $K \coloneqq \frac{1}{R_xR_y} = \frac{1 - \beta}{ {R_y}^2 }$. 

\eqnname{~\eqref{eqn: zero pressure stiffness}} holds for almost all spheroidal shells, except close to the limit of infinitely long, circular cylindrical shells ($\beta \to 1$).
For instance, unpressurized cylinders  have a finite linear indentation stiffness~\cite{dePablo2003} whereas our shallow-shell result predicts zero stiffness. 
The reason is that for such shells with low Gaussian curvature, the indentation responses at zero pressure are dominated by long-wavelength components much larger than the curvature radius $R_y$, and the shallow-shell approximations thus break down.
Nevertheless, as we will see in \secref{CYL. low validity}, shallow-shell theory becomes valid as the internal pressure rises because of the appearance of a new deformation length scale, so our approach remains useful up to $\beta = 1$ for internally pressurized shells. 

The fact that the zero-pressure stiffness of a double-curved shell depends on its local Gaussian curvature, in hindsight, is quite sensible; in fact, we might have guessed this dependence in the first place, without explicitly carrying out the integration. 
The reason is as follows. 
The physics should not depend on the choice of coordinates. 
In light of this, the zero-pressure stiffness can only depend on those quantities, constructed from the curvature tensor, that are invariant under rotation. 
For a two-dimensional surface, there are two such candidates, namely: the Gaussian curvature and the mean curvature. 
From \eqnref{eqn: Fourier stiffness}, we can infer that the stiffness simply cannot depend on the mean curvature: setting either $R_x$ or $R_y$ to infinity, the stiffness would vanish, while the mean curvature remains finite; in other words, it is, ironically, the inevitable failure of shallow-shell theory, when being applied to long cylindrical shells, that actually guarantees the Gaussian-curvature dependence. 
%%%%%%%%%%%%%%%%%%%%%%%%%%%%%%%%%%%%%%%%%%%%%%%%%%%%%%%%%%%%%%%%%%%%%%%%%%%%%%%%%%%%%%%%%%%%%%%%%%%%%%%%%%%%%%%%%%%%%%%%%%%%%%%%%%%%
%%%%%%%%%%%%%%%%%%%%%%%%%%%%%%%%%%%%%%%%%%%%%%%%%%%%%%%%%%%%%%%%%%%%%%%%%%%%%%%%%%%%%%%%%%%%%%%%%%%%%%%%%%%%%%%%%%%%%%%%%%%%%%%%%%%%
\subsection{Numerical Evaluation of the Stiffness Integral} 

\begin{figure}[ht] 
\centering 
\includegraphics[width = 0.48\textwidth]{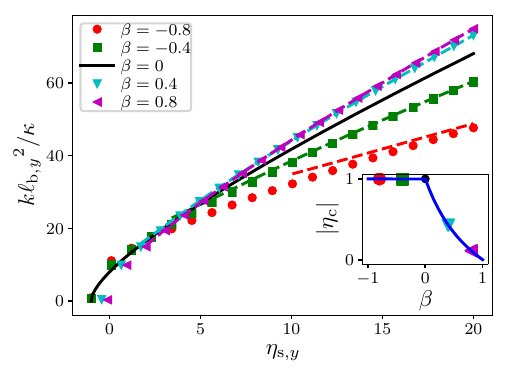} 
\caption{
Indentation stiffness of five different shell geometries subject to both internal and external pressures. 
Symbols denote data obtained from numerically evaluating the stiffness integral (\eqnname{~\eqref{eqn: 2D stiffness integral}}), and the dashed curves represent values associated with the analytical expression for the asymptotic indentation stiffness (\eqnname{~\eqref{eqn: asymptotic stiffness integral}}).
Solid line shows the known analytical stiffness for spherical shells, \eqnref{eqn: SPH. stiffness}.
Inset shows the magnitude of external pressures at which the indentation stiffness vanishes for each shell (symbols), compared to the prediction from the local instability criterion, \eqnname{~\eqref{eqn: ND. buckling pressure}} (solid line). 
} 
\label{fig: the summary} 
\end{figure} 

Having preliminarily verified the validity of the stiffness integral, we evaluate the indentation stiffness for four shell geometries in the range of interest $-1 < \beta \leq 1$ via numerical evaluation of the stiffness integral, \eqnref{eqn: 2D stiffness integral}. 
Results are shown in \figref{fig: the summary}.
As expected from \eqnref{eqn: zero pressure stiffness}, we observe that close to zero pressure, shells with lower values of $\beta$ are stiffer since their local Gaussian curvature is higher at the indentation point.
However, at higher pressures, the trend is reversed, and oblate shells become softer than prolate shells at the same pressure.
At negative (i.e., external) pressures, the indentation response softens, and the stiffness vanishes at a critical pressure value which falls with increasing asphericity for $\beta > 0$, but is constant at $\p = -1$ for $\beta < 0$ (inset to \figref{fig: the summary}). 
Below the critical pressure (i.e., at pressures more negative than the critical pressure), the indentation stiffness remains nil since the stiffness integral no longer converges; in fact, the indentation stiffness below the critical pressure, the concept itself becomes physically unmeaning because the shell is already buckled. 
In the remainder of this section, we reveal the physical mechanisms underlying these features through an analysis of the stiffness integral.
We also take a detailed look at the behavior of the indentation stiffness at pressures close to the critical value, and at high pressures.
We wrap up the section by studying the stiffness of pressurized cylinders ($\beta = 1$). 
%%%%%%%%%%%%%%%%%%%%%%%%%%%%%%%%%%%%%%%%%%%%%%%%%%%%%%%%%%%%%%%%%%%%%%%%%%%%%%%%%%%%%%%%%%%%%%%%%%%%%%%%%%%%%%%%%%%%%%%%%%%%%%%%%%%% 
%%%%%%%%%%%%%%%%%%%%%%%%%%%%%%%%%%%%%%%%%%%%%%%%%%%%%%%%%%%%%%%%%%%%%%%%%%%%%%%%%%%%%%%%%%%%%%%%%%%%%%%%%%%%%%%%%%%%%%%%%%%%%%%%%%%%
\subsection{Loss of stiffness and buckling instability} \label{buckling}

From \figurename{~\ref{fig: the summary}}, we notice that for all the chosen asphericities, there exists a critical external (i.e., negative) pressure at which the indentation stiffness vanishes.
This softening indicates a divergence of the stiffness integral, which occurs when $\tilde{k} \to 0$ for some value(s) of the wavevector $\mathbf{q}$ (\eqnname{~\eqref{eqn: Fourier stiffness}}) heralding the existence of an unstable mode at that wavevector.
The instability in the shell shape due to the divergent mode is a local manifestation of the buckling instability exhibited by curved shells under a uniform external pressure~\cite{LL_Elasticity}.
In practice, a minuscule indentation force applied when the shell is close to the buckling instability would generate a large, sudden inversion in the shell near the indentation point.
The description of this post-buckled shape with large deflections goes beyond the reach of shallow-shell theory, involving sudden, often catastrophic changes in the enclosed shell volume; however, the approach to the buckling threshold itself can be captured using linear stability analysis.
The relation between indentation response and buckling has been employed experimentally as a non-destructive means to determine the buckling threshold of thin shells~\cite{Schneider2017, Marthelot2017, Abbasi2021}. 

The \emph{local} critical pressure at the equator of the spheroid is the threshold $\cp$ at which the stiffness integral first becomes unbounded, i.e., 
$$
\lim_{\p \to \cp^+} I(\beta, \p) = +\infty. 
$$
This threshold is obtained by finding the global minima of $\tilde{k}(\mathbf{q})$ in wavevector space and identifying the pressure at which they hit zero, which gives: 
\refstepcounter{equation} \label{eqn: buckling-pressure expressions} 
\begin{equation} 
\cp 
= 
\begin{cases} 
-\frac{1 - \beta}{1 + \beta}, & \text{for}\ 0 \leq \beta < 1, \\[0.25 em] 
-1,                           & \text{for}\ -1 < \beta < 0, 
\end{cases} 
\label{eqn: ND. buckling pressure} 
\tag{\theequation, a} 
\end{equation} 
or 
\begin{equation} 
\CP \coloneqq \cp\, p_\text{sc} 
= 
\begin{cases} 
-\frac{ 4\sqrt{ {\kappa}Y } }{2R_xR_y - {R_y}^2}, & \text{for}\ 0 \leq \beta < 1, \\[0.5 em] 
-\frac{ 4\sqrt{ {\kappa}Y } }{ {R_y}^2 },         & \text{for}\ -1 < \beta < 0, 
\end{cases} 
\tag{\theequation, b} 
\label{eqn: buckling pressure} 
\end{equation} 
in real units. 
More compactly, we can write $\CP$, in terms of the equatorial Gaussian curvature $K$ and the asphericity $\beta$, as 
\begin{equation} 
\CP = -\frac{4\sqrt{ {\kappa}Y }K}{ 1 + |\beta| }, 
\tag{\theequation, c} 
\label{eqn: compact buckling pressure} 
\end{equation} 
for all $\abs{\beta} < 1$. 

We used the word ``local'' to emphasize the fact that the critical pressure we have identified only characterizes the loss of stability at the equator of the spheroidal shell.
Other regions of the shell have different local curvatures, and might experience loss of stability at different values of the external pressure.
The \emph{global} buckling pressure of the spheroidal shell corresponds to the smallest magnitude of external pressure at which a local instability arises somewhere on the shell.
Noting that regions of highest Gaussian curvature are locally the stiffest, and from symmetry considerations, we expect \eqnname{~\eqref{eqn: compact buckling pressure}} to be the global buckling pressure for prolate shells ($\beta > 0$) for which the Gaussian curvature is lowest for points along the equator.
For oblate shells, by contrast, the Gaussian curvature is lowest at the two poles ($( {\pm}a, c, c )$ in \figurename{~\ref{fig: an ellipsoid}}), where the local geometry is spherical with radius $R_\text{p} = \frac{b^2}{a} = \frac{1 - \beta}{ \sqrt{K}}$, and the corresponding buckling pressure is 
\begin{equation} 
\CPv = -\frac{4\sqrt{ {\kappa}Y }}{ R_\text{p}^2 }= -\frac{4\sqrt{ {\kappa}Y }K}{ (1 - \beta)^2 }. 
\tag{\theequation, d} 
\label{eqn: buckling pressure at the vertices}
\end{equation}
As expected, $\abs{\CPv} < \abs{\CP}$ for oblate shells with $\beta < 0$.
The expressions \eqref{eqn: compact buckling pressure} for $0 \leq \beta < 1$ and \eqref{eqn: buckling pressure at the vertices} for $-1 < \beta < 0$ reproduce known results for the global buckling pressures of ellipsoidal shells~\cite{Russian_Review}.

Moving forward, we only consider pressures above the local buckling pressure at the indentation point (i.e. $p > \CP$) when evaluating the indentation stiffness.
Note that for oblate shells, this range includes external pressures for which the poles of the spheroid are past their buckling threshold.
However, our local-stiffness results are still useful for shells which match the elasticity and geometry of \figref{fig: an ellipsoid} locally in the vicinity of the equator but deviate from it further away (e.g., an oblate spheroid reinforced at the poles to prevent buckling).
When $\p > \cp$, the integrand in \eqnref{eqn: 2D stiffness integral} is guaranteed to be positive definite, and the integration over the radial coordinate can be carried out to leave behind a single integral: 
\begin{widetext} 
\begin{equation} 
I(\beta, \p) = \frac{1}{8\pi^2}\Iy\int_0^{2\pi} \mathrm{d}{\theta}\, 
               \frac 
               {i\frac{\pi}{2} 
             + \tanh^{-1}\left[ 
               \frac 
               { \p\left(1 + \beta\sin^2{\theta}\right) } 
               { \sqrt{\p^2\left(1 + \beta\sin^2{\theta}\right)^2 - \left(1 - \beta\sin^2{\theta}\right)^2} } 
               \right] 
               } 
               { \sqrt{\p^2\left(1 + \beta\sin^2{\theta}\right)^2 - \left(1 - \beta\sin^2{\theta}\right)^2} }. 
\label{eqn: 1D stiffness integral} 
\end{equation} 
\end{widetext} 
%%%%%%%%%%%%%%%%%%%%%%%%%%%%%%%%%%%%%%%%%%%%%%%%%%%%%%%%%%%%%%%%%%%%%%%%%%%%%%%%%%%%%%%%%%%%%%%%%%%%%%%%%%%%%%%%%%%%%%%%%%%%%%%%%%%%
%%%%%%%%%%%%%%%%%%%%%%%%%%%%%%%%%%%%%%%%%%%%%%%%%%%%%%%%%%%%%%%%%%%%%%%%%%%%%%%%%%%%%%%%%%%%%%%%%%%%%%%%%%%%%%%%%%%%%%%%%%%%%%%%%%%%
\subsection{Analytical Results for Spherical Shells} \label{spherical shells} 
In this section, we will recover and review some results for the indentation stiffness of spherical shells in the literature. 

Setting $\beta = 0$ (or, equivalently, $R_x = R_y$) in \eqnsname{~\eqref{eqn: buckling-pressure expressions}}, we first recover the critical pressure of spherical shells~\cite{zoelly1915ueber}, 
\refstepcounter{equation} \label{eqn: SPH.-buckling-pressure expressions} 
\begin{equation} 
p_\text{c, sph} \coloneqq \CP(R_x = R_y) 
                        = -\frac{ 4\sqrt{ {\kappa}Y } }{ {R_y}^2 }, 
\tag{\theequation, a} 
\label{eqn: SPH. buckling pressure} 
\end{equation} 
or, in the scaled units, 
\begin{equation} 
\eta_\text{c, sph} \coloneqq \cp(\beta = 0) 
                           = -1. 
\tag{\theequation, b} 
\label{eqn: ND. SPH. buckling pressure} 
\end{equation} 
(Recall that in our convention, $\p < 0$ corresponds to an external pressure.) 

The stiffness integral (\eqnname{~\eqref{eqn: 1D stiffness integral}}) also gets greatly simplified because the angular dependence vanishes when $\beta = 0$: 
$$
I(\beta = 0, \p) = \frac{1}{4\pi}\Iy 
                   \frac 
                   { i\frac{\pi}{2} + \tanh^{-1}\left(\frac{\p}{ \sqrt{\p^2 - 1} }\right) } 
                   { \sqrt{\p^2 - 1} }. 
$$
Taking the inverse and rewriting the resulting expression in terms of real physical quantities, we obtain the established result of the indentation stiffness of pressurized spherical shells~\cite{Vella2012a, Paulose2012}, 
\begin{align} 
\begin{split} 
\sk(p) & = \frac{ 8\sqrt{ {\kappa}Y } }{R_y} 
           \\[0.25 em] 
       & \ \ \times 
           \begin{cases} 
           \frac{ \sqrt{1 - \p^2} }{ 1 - \frac{2}{\pi}\arcsin{\p} }, 
           & \text{for}\ \abs{\p} < 1, 
           \\[0.5 em] 
           \pi\frac{ \sqrt{\p^2 - 1} }{ \ln\left(\frac{ \p + \sqrt{\p^2 - 1} }{ \p - \sqrt{\p^2 - 1} }\right) }, 
           & \text{for}\ \p \geq 1. 
           \end{cases} 
\end{split} 
\label{eqn: SPH. stiffness} 
\end{align} 

In practice, two limits of the indentation stiffness are of particular interest: the asymptotic behavior in the large-pressure limit $\p \gg 1$ and the critical behavior as the buckling pressure is approached (the limit $\p \to \eta_\text{c, sph}^+$). 
For the large-pressure limit, it can be shown that 
\begin{equation} 
\sk \sim \frac{ 4\pi\sqrt{ {\kappa}Y } }{R_y}\frac{\p}{ \ln{2\p} }. 
\label{eqn: SPH. asymptotic behavior} 
\end{equation} 
By expanding $\sk$ around $p_\text{c, sph}$, one can demonstrate that the indentation stiffness of spherical shells near the critical pressure scales as $\sqrt{ \frac{p_\text{c, sph} - p}{ p_\text{c, sph} } } = \sqrt{ 1 - \frac{\p}{\cp} }$. 

In the next section, after briefly explaining why the two limits are interesting, we will derive similar expressions for spheroidal shells. 
%%%%%%%%%%%%%%%%%%%%%%%%%%%%%%%%%%%%%%%%%%%%%%%%%%%%%%%%%%%%%%%%%%%%%%%%%%%%%%%%%%%%%%%%%%%%%%%%%%%%%%%%%%%%%%%%%%%%%%%%%%%%%%%%%%%%
%%%%%%%%%%%%%%%%%%%%%%%%%%%%%%%%%%%%%%%%%%%%%%%%%%%%%%%%%%%%%%%%%%%%%%%%%%%%%%%%%%%%%%%%%%%%%%%%%%%%%%%%%%%%%%%%%%%%%%%%%%%%%%%%%%%%
\subsection{Analytical Expressions for Indentation Stiffness of Spheroidal Shells at Low and High Pressures} \label{two behaviors} 
While the stiffness integral can be numerically integrated to obtain the indentation stiffness at any pressure and geometry within our prescribed limits, the analytical behavior of the stiffness at large pressures ($\p \gg 1$) and close to the buckling instability ($\p \to \cp^+$) is of special interest. 
The large-pressure regime is relevant to the mechanics of biological cells, which are often investigated with indentation assays~\cite{Arnoldi2000,Smith2000,Ivanovska2004,Deng2011}, using shell elasticity as a minimal model of the expected response.
The crowded internal environment of living cells leads to very high turgor pressures; some typical values that have been reported are 30 kPa for the bacterium \emph{Escherichia coli} (which corresponds to $\p \approx 15$)~\cite{Deng2011}, 2 MPa ($\p \sim 10^3$) for \emph{Bacillus subtilis}~\cite{Whatmore1990} and 2 MPa ($\p \approx 10$) for \emph{Saccharomyces cerevisiae} yeast cells~\cite{Smith2000}.
Analytical expressions for the asymptotic behavior of the indentation stiffness at large rescaled pressures will be useful to infer elastic properties, which themselves are sensitive to biological processes, by performing indentation measurements. 
Secondly,  shell buckling can be interpreted as a first-order phase transition with pressure as the order parameter~\cite{Paulose_Soft_Spots}, which ought to leave a signature in the indentation response.
Studying the limit $\p \to \cp^+$ will provide us with insights regarding the essence of the non-analyticity of the indentation stiffness near the critical pressure. 
%%%%%%%%%%%%%%%%%%%%%%%%%%%%%%%%%%%%%%%%%%%%%%%%%%%%%%%%%%%%%%%%%%%%%%%%%%%%%%%%%%%%%%%%%%%%%%%%%%%%%%%%%%%%%%%%%%%%%%%%%%%%%%%%%%%%
\subsubsection{The Large-Pressure Regime} \label{asymptotic} 
We are interested in finding a parameter in the modified stiffness integral, \eqnref{eqn: 1D stiffness integral}, which becomes small at large pressures to enable an exact evaluation of the leading stiffness behavior.
From the form of the integrand, the appropriate parameter is identified as 
$$
y(\beta, \p, \theta) \coloneqq \frac{1}{\p}\frac{ 1 - \beta\sin^2{\theta} }{ 1 + \beta\sin^2{\theta} }. 
$$
In the range of interest of the asphericity, $-1 < \beta \leq 1$, the parameter $y$ is small for all $0 \leq \theta < 2\pi$ provided the pressure satisfies  
\begin{equation} 
\p = \frac{p{R_y}^2}{ 4\sqrt{ {\kappa}Y } } 
   \gg 
     \begin{cases} 
     1,                           & \text{for}\ 0 \leq \beta \leq 1, \\[0.25 em] 
     \frac{1 - \beta}{1 + \beta}, & \text{for}\ -1 < \beta < 0. 
     \end{cases} 
\tag{$*$} 
\label{eqn: large-pressure criterion} 
\end{equation}
We use the small parameter to rewrite and analytically evaluate the modified stiffness integral (\eqnname{~\eqref{eqn: 1D stiffness integral}}) in the large-pressure limit: 
\begin{widetext} 
\begin{align} 
\begin{split} 
I & = \frac{1}{8\pi^2}\Iy\int_0^{2\pi} 
      \frac{ \mathrm{d}{\theta} }{ \p\left(1 + \beta\sin^2{\theta}\right) } 
      \frac{ i\frac{\pi}{2} + \tanh^{-1}\left(\frac{1}{ \sqrt{1 - y^2} }\right) }{ \sqrt{1 - y^2} } 
      \\[0.5 em] 
  & \stackrel{y \ll 1}{\approx} 
      \frac{1}{8\pi^2}\Iy\frac{1}{\p} 
      \left[ 
      \ln{2\p}\int_0^{2\pi} \frac{ \mathrm{d}{\theta} }{ 1 + \beta\sin^2{\theta} } 
    + \int_0^{2\pi} 
      \frac{ \mathrm{d}{\theta} }{ 1 + \beta\sin^2{\theta} } 
      \ln\left(\frac{ 1 + \beta\sin^2{\theta} }{ 1 - \beta\sin^2{\theta} }\right) 
      \right] 
      \\[0.5 em] 
  & \approx 
      \frac{1}{4\pi}\Iy\frac{1}{\p}\frac{1}{ \sqrt{1 + \beta} } 
      \left[ 
      \ln{4\p} 
    + \ln\left(1 + \frac{1}{\beta}\right) 
    - 2\tanh^{-1}\left(\sqrt{ \frac{1 - \beta}{1 + \beta} }\,\right) 
      \right]. 
\end{split} 
\label{eqn: asymptotic stiffness integral} 
\end{align} 
\end{widetext} 
The inverse of \eqnref{eqn: asymptotic stiffness integral} provides an analytical expression for the indentation stiffness of spheroidal shells at large pressures. We note that in our choice of length and pressure units, \eqnname{~\eqref{eqn: asymptotic stiffness integral}} holds for both prolate and oblate spheroidal shells, but the criterion for ``large pressure'' differs in these two cases, as defined in \eqnname{~\eqref{eqn: large-pressure criterion}}. 

To simplify and shed light on the final expression, we introduce a novel radius parameter $\mathcal{R} \coloneqq R_y\sqrt{1 + \beta}$ with which we can rewrite \eqnname{~\eqref{eqn: asymptotic stiffness integral}} as 
\refstepcounter{equation} \label{eqn: asymptotic novel stiffness integral} 
\begin{align} 
I & = \frac{1}{4\pi}\IR\frac{1}{\pR} 
      \left[\ln{4\pR} - \ln\left(1 + \sqrt{1 - \beta^2}\right)\right] 
      \tag{\theequation, a} 
      \label{eqn: asymptotic novel stiffness integral 1} 
      \\[0.25 em] 
  & = \frac{1}{4\pi}\IR\frac{1}{\pR} 
      \left[\ln{4\pR} - \ln\left(1 + \mathcal{R}\sqrt{K}\right)\right], 
      \tag{\theequation, b} 
      \label{eqn: asymptotic novel stiffness integral 2} 
\end{align} 
where 
$
\pR \coloneqq \frac{p\mathcal{R}^2}{ 4\sqrt{ {\kappa}Y } } 
            = \p(1 + \beta) 
$
is the corresponding scaled pressure, and $K$ again denotes the local Gaussian curvature of the given shell. 
Besides its mathematical convenience, $\mathcal{R}$ can be related to the second stress invariant, i.e., the determinant of the prestress tensor $\sigma_{\alpha\beta}^0$: 
\begin{equation} 
\label{eqn:novel_R} 
\mathcal{R} = R_y\sqrt{1 + \beta} 
            = R_y\sqrt{ 2 - \frac{R_y}{R_x} } 
            = \frac{2}{p}\sqrt{\sigma_{11}^0\sigma_{22}^0}, 
\end{equation} 
where $\sigma_{11}^0 = \partial_{22}{\chi_0}$, and $\sigma_{22}^0 = \partial_{11}{\chi_0}$ are the prestresses along the principal directions in the membrane state.
Equation~\ref{eqn:novel_R} shows that $\mathcal{R}$ is the radius of curvature for which the internal pressure $p$ would balance a membrane tension of magnitude equal to the square root of the second stress invariant according to Laplace's law~\cite{deGennes2013}. We term the associated curvature, $\mathcal{R}^{-1}$, the \emph{distensile curvature}.
Just as the local Gaussian curvature dictates the zero-pressure indentation stiffness of curved shells (\eqnref{eqn: zero pressure stiffness}), the distensile curvature dominates the indentation response at large internal pressures although a residual dependence on the Gaussian curvature remains in the stiffness integral (\eqnref{eqn: asymptotic novel stiffness integral 2}).

\begin{figure}[htb] 
\centering 
\includegraphics[width = 0.48 \textwidth]{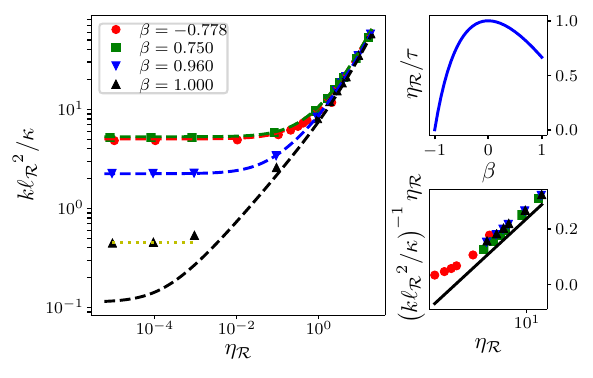} 
\caption{
  Comparison of predicted indentation stiffness (dashed lines) to finite-element simulation data from Ref.~\onlinecite{Vella2012} (symbols).  Data are scaled using the length scale  $\ell_\mathcal{R} = \sqrt[4]{ \frac{ {\kappa}{ \mathcal{R} }^2 }{Y} }$ and pressure scale $4\sqrt{\kappa Y}/\mathcal{R}^2$, for which we predict convergence of the stiffness curves at large rescaled pressures $\pR = p\mathcal{R}^2/4\sqrt{\kappa Y}$.
  A similar convergence was depicted in FIG.~3 of Ref.~\onlinecite{Vella2012} using different scales.
  Yellow solid line corresponds to the zero-pressure stiffness of long cylindrical shells, calculated with formulae in Ref.~\onlinecite{dePablo2003}, since shallow-shell theory does not apply to cylinders below a threshold pressure (see discussion in \secref{cylindrical shells}).
  The top inset compares the rescaled pressure $\pR$ to the alternative variable $\tau$ introduced in Ref.~\onlinecite{Vella2012}. (See text for details.) 
  The bottom inset is a linear-log plot showing how the product of the inverse scaled stiffness and the scaled pressure varies as the pressure increases; thereinto, the black solid line corresponds to the result for spherical shells. 
} 
\label{fig: Vella} 
\end{figure}

From \eqnname{~\eqref{eqn: asymptotic novel stiffness integral 1}} we can see that under the rescaling (with the novel radius parameter), spherical shells are the stiffest in the large-pressure regime since when $\beta = 0$, the geometric contribution, $\ln\left(1 + \sqrt{1 - \beta^2}\right)$, reaches its maximum $\ln{2}$ and hence minimizes the stiffness integral. (This feature is depicted in the bottom inset of \figref{fig: Vella}.) 
At still higher pressures such that the geometric contribution becomes negligible, \eqnname{~\eqref{eqn: asymptotic novel stiffness integral 1}} reduces to 
\begin{equation} 
I \stackrel{\pR \gg 1}{\approx} \frac{1}{4\pi}\IR\frac{ \ln{4\pR} }{\pR}. 
\label{eqn: asymptotic scaling form} 
\end{equation} 
In practice, the expression 
\begin{equation} 
I \stackrel{\pR \gg 1}{\approx} \frac{1}{4\pi}\IR\frac{ \ln{2\pR} }{\pR}
\label{eqn: asymptotic scaling form 2} 
\end{equation} 
is more accurate for most shell geometries, since $\ln\left(1 + \sqrt{1 - \beta^2}\right)$ is closer to $\ln{2}$ than to zero for $|\beta| < 0.91$.
By comparison to \eqnref{eqn: SPH. asymptotic behavior}, we see that the corresponding stiffness is identical to the high-pressure response of a spherical shell with radius $\mathcal{R}$ and rescaled pressure $\pR$. 

Equations~\eqref{eqn: asymptotic novel stiffness integral} and \eqref{eqn: asymptotic scaling form 2} provide a concise interpretation of the indentation stiffness of spheroidal shells at large pressures.
When the material parameters $\kappa$ and $Y$ are fixed, the stiffness in the large-pressure limit depends on three quantities: the pressure $p$, the determinant of the stress tensor at the point of indentation, and the local Gaussian curvature $K$.
The first two quantities define a curvature radius $\mathcal{R}$ and a dimensionless pressure $\pR$ which both originate from the membrane prestress; \eqnref{eqn: asymptotic novel stiffness integral 2} explicitly separates the prestress and geometry contributions to the indentation stiffness.
The large-pressure indentation of the shell approaches that of a sphere with the prestress-derived curvature and pressure scales (\eqnref{eqn: asymptotic scaling form 2}) when the weak dependence on $K$ is ignored.
Upon using these new scales, a duality connecting prolate to oblate shells at high pressures is revealed: \eqnref{eqn: asymptotic novel stiffness integral 1} is invariant under the replacement $\beta \to -\beta$, so a shell with geometric parameters $\left\{R_y = \rho,\ \beta = \beta_0\right\}$ and internal pressure $p$ has the same high-pressure response as a shell with parameters $\left\{R_y = \rho\sqrt{ (1 + \beta_0) / (1 - \beta_0) },\ \beta = -\beta_0\right\}$ and the same pressure, for which the parameters $\mathcal{R}$ and $\pR$ are identical.
The criterion for high pressure, \eqnref{eqn: large-pressure criterion}, also reduces to the symmetric form $$\pR \gg 1+ |\beta| \sim 1.$$

\paragraph{Comparison with Established Results.} 
In previous works~\cite{Lazarus2012a,Vella2012}, it was hypothesized that the high-pressure indentation response of ellipsoidal shells is dictated by the mean curvature radius $\Rm = 2/(R_x^{-1}+R_y^{-1})$ and a dimensionless pressure scale $\tau$  set by the mean membrane prestress at the indentation point, $\sigmam = (\sigma_{11}^0+\sigma_{22}^0)/2$, via
$$\tau = \frac{\sigmam  \Rm}{2 \sqrt{\kappa Y}}.$$
Our asymptotic form for the inverse of the indentation stiffness, \eqnref{eqn: asymptotic scaling form 2}, parallels the high-pressure indentation stiffness proposed in Ref.~\onlinecite{Vella2012} which used $\Rm$ and $\tau$ in place of $\mathcal{R}$ and $\pR$ respectively.
However, the origins of our radius and pressure scales are somewhat different as they utilize the determinant, rather than the trace, of the membrane stress (\eqnref{eqn:novel_R}).

Despite these differences, our proposed length and pressure scales are as successful as the previously-hypothesized scales in quantifying the high-pressure indentation response.
In \figref{fig: Vella}, we compare our predictions for the indentation stiffness (curves) to the results of finite-element simulations (symbols) reported in FIG.~3 of Ref.~\onlinecite{Vella2012}.
Upon using the new length and pressure scales, the data for $\pR \gg 1$ collapse onto the proposed asymptotic form, \eqnref{eqn: asymptotic scaling form 2} (solid line).
We also compare our numerically-evaluated indentation stiffness, \eqnref{eqn: 1D stiffness integral} (dashed curves), to the finite-element data, and find quantitative agreement for almost all geometries and pressures.
The disagreement between the stiffness integral and the measured indentation stiffness at low pressures for cylindrical shells ($\beta = 1$) is expected; see \secref{cylindrical shells} for an explanation and a more accurate prediction (dotted line).
Our prediction also deviates from the finite-element simulation results for the highest simulated pressure of the oblate spheroidal geometry $\beta = -0.778$; we hypothesize that second-order shape changes in response to the internal pressure might be responsible for this discrepancy.
Apart from these data points, our theoretical predictions lie within 5\% of the finite-element measurements, which  validates our approach over a wide range of geometries and pressures.
The duality connecting prolate to oblate shells upon using the scales $\mathcal{R}$ and $\pR$ is also visible in \figref{fig: Vella}, since the data for $\beta = 0.75$ and $\beta = -0.778$ nearly overlap. 

We were unable to neatly separate the contributions from the mean and the Gaussian curvatures in our evaluation of the high-pressure stiffness integral.
Therefore, we cannot directly evaluate the relative merits of using our proposed scales $\mathcal{R}$ and $\pR$ over the previously-proposed scales $\Rm$ and $\tau$ from Ref.~\onlinecite{Vella2012}.
However, some insight as to why both scales perform well in explaining the high-pressure indentation stiffness can be obtained by comparing them as a function of geometry. 
By expressing $\Rm$ and  $\tau$ in terms of $R_y$ and $\beta$, we find the ratio 
$$
\frac{\pR}{\tau} = \frac{ (1 + \beta)(2 - \beta) }{2 + \beta}. 
$$
As the inset of \figurename{~\ref{fig: Vella}} illustrates, the ratio is of order one for most values of the asphericity.
Similarly, the ratio $\mathcal{R}/\Rm$ evaluates to a number of order one for $|\beta| < 1$.
Therefore, using the two sets of physical scales is expected to provide similar results.
The discrepancy between the two approaches becomes significant only for oblate shells with $\beta$ approaching $-1$.
In this limit, an advantage of the scales introduced here is that the instability  expected for large internal pressures at $\beta = -1$ (see \secref{buckling}) is reflected in the pressure-induced curvature taking on imaginary values when $\sigma_{22}^0$ becomes negative in \eqnref{eqn:novel_R}.
By contrast, the mean curvature and mean prestress both remain positive and vary smoothly as $\beta$ falls below $-1$, and the approximations using the mean scales incorrectly predict a finite indentation response at large pressures. 
%%%%%%%%%%%%%%%%%%%%%%%%%%%%%%%%%%%%%%%%%%%%%%%%%%%%%%%%%%%%%%%%%%%%%%%%%%%%%%%%%%%%%%%%%%%%%%%%%%%%%%%%%%%%%%%%%%%%%%%%%%%%%%%%%%%%
\subsubsection{The Critical Behavior upon Approaching the Buckling Pressure} \label{critical} 

\begin{figure}
\centering 
\includegraphics[width = 0.48\textwidth]{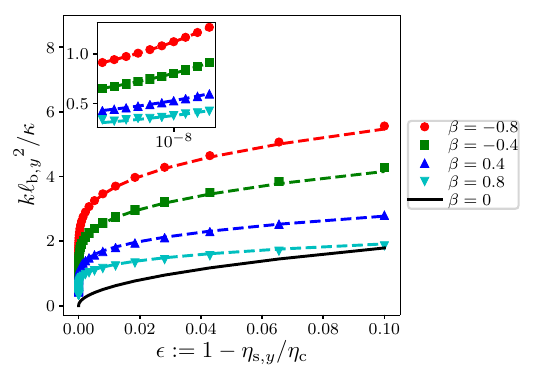} 
\caption{
Critical behavior of the indentation stiffness as $\cp$ is approached, for five spheroidal geometries. 
Symbols denote data obtained from numerical integration of the stiffness integral. 
Solid and dashed lines correspond to analytical expressions \eqnsname{~\eqref{eqn: SPH. stiffness}} and \eqref{eqn: critical stiffness integral}, respectively. 
Inset shows the same data on linear-log scales to reveal the slow approach to zero stiffness as $\p \to \cp^+$ for non-spherical shells. 
} 
\label{fig: the critical} 
\end{figure}

We now analyze the functional approach of the indentation stiffness to zero as the critical pressure is approached from above ($\p \to \cp^+$).
(Recall that in our convention, external pressures correspond to $\p < 0$, and the critical pressure $\cp$ for the local instability is negative.) 
Defining the fractional distance from the critical pressure as 
\begin{align*} 
\epsilon & \coloneqq \frac{\cp - \p}{\cp} 
                     \\[0.25 em] 
         &         = \begin{cases} 
                     1 + \frac{1 + \beta}{1 - \beta}\p, & \text{for}\ 0 \leq \beta < 1, \\[0.25 em] 
                     1 + \p,                            & \text{for}\ -1 < \beta < 0, 
                     \end{cases} 
\end{align*} 
we would like to study the limit $\epsilon \to 0^+$ of the stiffness integral. 
As before, we first rewrite the stiffness integral (\eqnname{~\eqref{eqn: 1D stiffness integral}}) in a simplified form 
\begin{align*} 
I & = \frac{1}{2\pi^2}\Iy\int_0^\frac{\pi}{2} \mathrm{d}{\theta}\, 
      \\[0.25 em] 
  & \ \ \times 
      \begin{cases} 
      \frac{1}{ 1 - \beta\cos^2{\theta} } 
      \frac{ \frac{\pi}{2} + \arcsin{y} }{ \sqrt{1 - y^2} }, 
      & \text{for}\ 0 \leq \beta < 1, 
      \\[1.25 em] 
      \frac{1}{ 1 - \beta\sin^2{\theta} } 
      \frac{ \frac{\pi}{2} + \arcsin{y'} }{ \sqrt{1 - y'^2} }, 
      & \text{for}\ -1 < \beta < 0, 
      \end{cases} 
\end{align*} 
where 
$$
y(\beta, \epsilon, \theta) \coloneqq \frac{ 1 + \beta\cos^2{\theta} }{1 + \beta}\frac{1 - \beta}{ 1 - \beta\cos^2{\theta} }(1 - \epsilon), 
$$
and 
$$
y'(\beta, \epsilon, \theta) \coloneqq \frac{ 1 + \beta\sin^2{\theta} }{ 1 - \beta\sin^2{\theta} }(1 - \epsilon). 
$$
Notice that the limits 
$$
\lim_{ \substack{\epsilon \to 0^+ \\[0.1 em] \theta \to 0^+} } y = 1 
\quad \text{and} \quad 
\lim_{ \substack{\epsilon \to 0^+ \\[0.1 em] \theta \to 0^+} } y' = 1 
$$
give rise to divergence of the integrals. 
In other words, in the limit $\epsilon \to 0^+$, the definite integrals are dominated by their values in the vicinity of $\theta = 0$. 
Accordingly, we can approximate them by replacing the integrands with the corresponding second-order Taylor polynomials around $(\epsilon, \theta) = (0, 0)$: 
\begin{align*} 
I & \approx \frac{1}{2\sqrt{2}\pi}\Iy\int_0^\frac{\pi}{2} \mathrm{d}{\theta}\, 
            \\[0.25 em] 
  & \ \ \times 
            \begin{cases} 
            \frac{1}{ 1 - \beta(1 - \theta^2) } 
            \frac{1}{ \sqrt{ \epsilon + \frac{2\beta\theta^2}{1 - \beta^2} } }, 
            & \text{for}\ 0 \leq \beta < 1, 
            \\[1.25 em] 
            \frac{1}{ (1 - \beta\theta^2)\sqrt{\epsilon - 2\beta\theta^2} }, 
            & \text{for}\ -1 < \beta < 0. 
            \end{cases} 
\end{align*} 
These integrals can be analytically evaluated: 
\begin{widetext} 
\begin{equation} 
I \approx \frac{1}{4\pi}\Iy\frac{1}{ \sqrt{ \abs{\beta\cp} } } 
          \sinh^{-1}\left( 
          \sqrt{ \frac{2}{ \frac{4}{\pi^2}\frac{1}{ \abs{\beta} } + 1 + f(\beta) } }\frac{1}{ \sqrt{\epsilon} } 
          \right) 
  \approx \frac{1}{4\pi}\Iy\frac{1}{ \sqrt{ \abs{\beta\cp} } } 
          \ln\left( 
          \sqrt{ \frac{8}{ \frac{4}{\pi^2}\frac{1}{ \abs{\beta} } + 1 + f(\beta) } }\frac{1}{ \sqrt{\epsilon} } 
          \right), 
\label{eqn: critical stiffness integral} 
\end{equation} 
\end{widetext} 
where 
$$
f(\beta) \coloneqq \begin{cases} 
                   \left(1 - \frac{4}{\pi^2}\right)\beta, & 
                   \text{for}\ 0 \leq \beta < 1, \\[0.25 em] 
                   0, & 
                   \text{for}\ -1 < \beta < 0. 
                   \end{cases} 
$$
\figref{fig: the critical} shows that the analytical result, \eqnref{eqn: critical stiffness integral} (dashed lines) successfully reproduces the results due to numerical integration of the stiffness integral (symbols) close to the critical pressure for four different shells. 

\eqnname{~\eqref{eqn: critical stiffness integral}}, the main result in this section, implies that for general spheroidal shells, their indentation stiffness falls off as the inverse of the logarithm of the distance $\epsilon$ from the critical point, i.e., $k \propto -\frac{1}{ \ln{ \sqrt{\epsilon} } }$.
This slow approach of the stiffness to zero, due to the logarithmic divergence of the stiffness integral for non-spherical shells, is evident in the inset to \figref{fig: the critical}.
By contrast, the approach to zero is more drastic for spherical shells: upon taking the limit $|\beta| \to 0$ of \eqnref{eqn: critical stiffness integral}, one obtains 
$$
  \lim_{ |\beta| \to 0 } \frac{1}{ \sqrt{ |\beta\cp| } } 
  \sinh^{-1}\left( 
  \sqrt{ \frac{2}{\frac{4}{\pi^2}\frac{1}{ |\beta| } + 1} }\frac{1}{ \sqrt{\epsilon} } 
  \right) 
= \frac{\pi}{ \sqrt{2\epsilon} }, 
$$
which implies that near the critical pressure, the indentation stiffness of spherical shells $\sk \propto \sqrt{\epsilon}$, as was expected from the exact results reported in \secref{spherical shells}.
Spherical shells are much softer at all pressures near the critical pressure compared to spheroidal shells, since $\sk / k \propto \sqrt{\epsilon}\ln{ \sqrt{\epsilon} } \stackrel{\epsilon \to 0}{\to} = 0$.
The contrasting characters of the softening as the critical pressure is approached reflects the fact that spherical shells harbor a massive degeneracy of divergent Fourier components of the stiffness integral as $\p \to \cp^+$ (a circle with radius $q = 1 / \elly$ in the wavevector plane), whereas non-spherical shells exhibit divergent Fourier modes only at the two values, $\mathbf{q}_\pm^\text{pro} = \left(0, \pm \sqrt{1 - \beta} / \elly\right)$ and $\mathbf{q}_\pm^\text{ob} = (\pm 1 / \elly, 0)$ for prolate and oblate shells, respectively.

The results in this subsection do not apply to cylindrical shells because the limit $\beta \to 1^-$ fails to exist ($\lim_{\beta \to 1^-} \cp = 0$). 
A different approach will be used to study cylindrical shells below.
%%%%%%%%%%%%%%%%%%%%%%%%%%%%%%%%%%%%%%%%%%%%%%%%%%%%%%%%%%%%%%%%%%%%%%%%%%%%%%%%%%%%%%%%%%%%%%%%%%%%%%%%%%%%%%%%%%%%%%%%%%%%%%%%%%%%
%%%%%%%%%%%%%%%%%%%%%%%%%%%%%%%%%%%%%%%%%%%%%%%%%%%%%%%%%%%%%%%%%%%%%%%%%%%%%%%%%%%%%%%%%%%%%%%%%%%%%%%%%%%%%%%%%%%%%%%%%%%%%%%%%%%%
\subsection{Indentation Responses of Cylindrical Shells} \label{cylindrical shells} 
The case of extremely long prolate ellipsoids, for which $R_x \to \infty$ or $\beta \to 1$, requires special treatment.
As we argued in \secref{methods}, shallow-shell theory accurately describes the strains associated with transverse deflections only if the deflections vary over length scales that are small compared to the local radii of curvature $R_x$ and $R_y$.
When $R_x \to \infty$, the characteristic wavelength of deflections along the $y$ direction, which is controlled by $\ellx$, eventually becomes larger than the cylinder circumference $2\pi R_y$.
Shallow-shell theory builds the response to point indentation and to external pressure out of modes that do not change the metric or curvature of the shell, erroneously predicting zero indentation stiffness for unpressurized cylinders (\secref{zero pressure}) as well as buckling at an infinitesimal external pressure ($\cp \to 0$, \secref{buckling}).

The true deformation mode responsible both for the indentation stiffness and the finite buckling pressure of an infinitely long cylindrical shell is the isometric change in shape of the circular cross section to an ellipse, which is also responsible for the buckling of inextensible rings~\cite{Ventsel2002}.
This mode extends over the entire shell circumference and cannot be captured by shallow-shell theory.
Unlike the characteristic deflections of doubly-curved shell segments which involve both stretching and bending, the elliptical mode costs no stretching energy as it does not change the circumference; it only involves bending energy because of the change in curvature away from the initial circular shape.
When the bending energy of the elliptical shape change is evaluated using basis functions that extend over the entire circumference, the mechanics of the cylindrical shell can be accurately described.
For instance, the buckling pressure for long cylindrical shells evaluated using elliptical modes is $p_\text{c, cyl} = -\frac{3\kappa}{ {R_y}^3 }$~\cite{Ventsel2002}. 
Using similar methods, the indentation stiffness of zero-pressure cylindrical shells was derived in Ref.~\onlinecite{dePablo2003} to be 
\begin{equation} 
\ckzero \approx 1.37\frac{ Et^\frac{5}{2} }{ R^\frac{3}{2} }. 
\label{eqn:dePablo} 
\end{equation} 
In principle, the same non-shallow-shell techniques could be used to evaluate the indentation stiffness of cylinders at finite internal pressure.
However, we find that the balance between pressure and elasticity gives rise to a characteristic deformation wavelength that quickly becomes small compared to the curvature radius as the internal pressure is increased, and the validity of shallow-shell results becomes re-established.
To derive this new characteristic length, we notice that it is bending and not stretching which primarily dictates the deformation energy of cylinders because of the existence of the isometric deformation modes, as is evident from the expressions for the buckling pressure and the zero-pressure indentation stiffness. 
Accordingly, the characteristic extent of deformations of pressurized cylinders is obtained by balancing the bending and tension terms in the total-energy functional, which leads to the result
$$
\ellp = \sqrt{ \frac{\kappa}{pR_y} }. 
$$
When the pressure becomes appreciable, this length scale falls far below the curvature radius $R_y$ and our shallow-shell analysis, in particular the stiffness integral \eqnref{eqn: 1D stiffness integral}, can be used to derive the indentation of cylindrical shells.
In \figref{fig: Vella}, the numerically-integrated indentation stiffness (black dashed line) is seen to agree with the results of finite-element simulations (from Ref.~\onlinecite{Vella2012}) for pressures above $\pR = 2 \p =0.1$.
At lower pressures, the stiffness crosses over to the zero-pressure result from Ref.~\onlinecite{dePablo2003} (yellow line).

In the remainder of this subsection, we develop analytical approximations for the indentation stiffness of cylinders which cover a wide range of pressures.
To do so, we exploit the separation between the two pressure scales $\p$ and $|p_\text{c, cyl}|$ for thin shells, which allows us to evaluate the stiffness integral in ``low-pressure'' ($\p \ll 1$) and ``high-pressure'' ($\p \gg 1$) regimes while satisfying the condition $\ellp \ll R_y$ for shallow-shell theory to be valid.
%%%%%%%%%%%%%%%%%%%%%%%%%%%%%%%%%%%%%%%%%%%%%%%%%%%%%%%%%%%%%%%%%%%%%%%%%%%%%%%%%%%%%%%%%%%%%%%%%%%%%%%%%%%%%%%%%%%%%%%%%%%%%%%%%%%%
\subsubsection{Validity of Shallow-Shell Theory in the Low-Pressure Limit $\p \ll 1$} \label{CYL. low validity} 
First, we will show that the shallow-shell theory is still valid for a range of pressures satisfying $\p \ll 1$ for long cylindrical shells, provided that the shells are thin. 
The requirement $\ellp \ll R_y$ amounts to $p \gg \frac{\kappa}{ {R_y}^3 }$, a pressure scale related to the buckling pressure  $p_\text{c, cyl}$.
To relate the two pressure scales $p_\text{sc}$ (which underlies the dimensionless pressure $\p$) and $\frac{\kappa}{ {R_y}^3 }$, we write: 
\begin{equation} 
\p      = \frac{p{R_y}^2}{ 4\sqrt{ {\kappa}Y } } 
   \equiv \frac{1}{4}\frac{p{R_y}^3}{\kappa}\frac{1}{ \sqrt{\gamma_y} }, 
\label{eqn: two pressure scales} 
\end{equation} 
where the dimensionless F\"{o}ppl-von K\'{a}rm\'{a}n number is 
$$
\gamma_y \equiv \frac{Y{R_y}^2}{\kappa} 
         \simeq 10\left(\frac{R_y}{t}\right)^2. 
$$
For typical thin elastic shells, $0.001 \leq \frac{t}{R_y} \leq 0.05$~\cite{Ventsel2002}, so there exists a wide range of pressures which simultaneously satisfy $p \gg \frac{\kappa}{ {R_y}^3 }$ (that is, $\ellp \ll R_y$, and the shallow-shell theory holds) and $\p \ll 1$ (the low-pressure limit for the stiffness integral). 
The wide range manifests the fact that due to a finite Gaussian curvature, a spherical shell is able to withstand more external forces than a long cylindrical shell of the same radius, i.e., the magnitude of the spherical shell's buckling pressure is larger. 
%%%%%%%%%%%%%%%%%%%%%%%%%%%%%%%%%%%%%%%%%%%%%%%%%%%%%%%%%%%%%%%%%%%%%%%%%%%%%%%%%%%%%%%%%%%%%%%%%%%%%%%%%%%%%%%%%%%%%%%%%%%%%%%%%%%%
\subsubsection{Analytical Expressions for the Indentation Stiffness in the Limit $\p \ll 1$} \label{CYL. low} 
We start our analysis of the stiffness integral (\eqnname{~\eqref{eqn: 1D stiffness integral}}) by setting $\beta = 1$: 
$$
I(\beta = 1, \p) = \frac{1}{2\pi^2}\Iy\int_0^\frac{\pi}{2} 
                   \frac{ \mathrm{d}{\theta} }{ \sin^2{\theta} } 
                   \frac{ \arccos\big(y(\p, \theta)\big) }{ \sqrt{ 1 - y^2(\p, \theta) } }, 
$$
where 
$$
y(\p, \theta) \coloneqq \p\frac{ 1 + \cos^2{\theta} }{ \sin^2{\theta} } 
                      = \p\left(2\cot^2{\theta} + 1\right). 
$$
Making the substitution $u = \cot{\theta}$, we can further reduce the stiffness integral: 
\begin{align*} 
I(\beta = 1, \p) & = \frac{1}{2\pi^2}\Iy\frac{1}{ \sqrt{2\p} } 
                     \\[0.25 em] 
                 & \ \ \times 
                     \int_0^{+\infty} \mathrm{d}{u}\, 
                     \frac{ \arccos(u^2 + \p) }{ \sqrt{1 - (u^2 + \p)^2} }. 
\end{align*} 
Note that $\p$ now couples with the integration variable $u$ in an additive manner. 
Hence, for $\p \ll 1$, we can write the stiffness integral as a power series of $\p$: 
\begin{widetext} 
\begin{equation} 
I = \frac{1}{8\pi^2}\Iy \times 2\sqrt{2} 
    \sum_{n = 0}^\infty \frac{1}{n!} 
    \left[\int_0^{+\infty} \mathrm{d}{u}\, \left(D_{u^2}\right)^n\left(\frac{ \arccos\left(u^2\right) }{ \sqrt{1 - u^4} }\right)\right] 
    \left(\p\right)^{ n - \frac{1}{2} }, 
\label{eqn: CYL. low} 
\end{equation} 
%\end{widetext} 
where the differential operator 
$
D_{u^2} \equiv \frac{ \mathrm{d} }{ \mathrm{d}(u^2) } 
             = \frac{1}{2u}\frac{ \mathrm{d} }{ \mathrm{d}{u} }. 
$
Truncating the series after the first four terms and numerically evaluating the coefficients gives the sought approximate expression for the indentation of cylindrical shells with pressures in the range $\kappa/R_y^3 \ll p \ll 4 \sqrt{\kappa Y}/R_y^2$:
%\begin{widetext} 
\begin{equation} 
I \approx \frac{1}{8\pi^2}\Iy 
          \left( 
          \frac{11.6}{ \sqrt{\p} } 
        - 2.66\sqrt{\p} 
        + 1.83\p^\frac{3}{2} 
        - 0.998\p^\frac{5}{2} 
          \right). 
\label{eqn: truncated CYL. low} 
\end{equation} 
\end{widetext} 
%%%%%%%%%%%%%%%%%%%%%%%%%%%%%%%%%%%%%%%%%%%%%%%%%%%%%%%%%%%%%%%%%%%%%%%%%%%%%%%%%%%%%%%%%%%%%%%%%%%%%%%%%%%%%%%%%%%%%%%%%%%%%%%%%%%%
\subsubsection{Analytical Expressions for the Indentation Stiffness of Highly-Pressurized Cylindrical Shells} \label{CYL. high} 

\begin{figure}[htb] 
\centering 
\includegraphics[width = 0.48\textwidth]{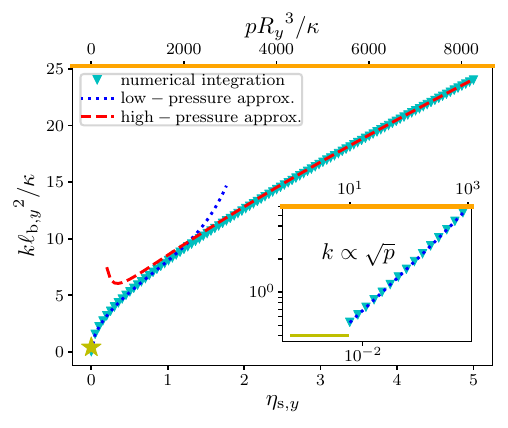} 
\caption{
  Scaled indentation stiffness as a function of pressure for an infinitely long, thin circular cylindrical shell.
  Results of numerical integration (symbols) are compared to the two approximate expressions derived for low pressures (dotted curve) and high pressures (dashed curve). 
  The lower axis reports the rescaled pressure $\p = p/p_\text{sc}$, whereas the upper axis (orange) uses the alternate pressure scale $\kappa/ {R_y}^3$ computed using the parameters $\kappa = 1.76 \times 10^{-19}\ \mathrm{J}$ and $R_y = 0.5\ \mathrm{ {\mu}m }$ representative of the \emph{E. coli} cell wall~\cite{Deng2011}.
The resulting \FvK number is $\gamma_y = 1.71 \times 10^5$. 
Shallow-shell theory is valid as long as $\frac{p{R_y}^3}{\kappa} \gtrsim 10$.
The zero-pressure stiffness $\ckzero$ (\eqnref{eqn:dePablo}), calculated using different methods in Ref.~\onlinecite{dePablo2003}, is marked by a star in the main panel and a solid line in the inset.
Inset shows the low-pressure behavior on logarithmic scales, where the square-root dependence of indentation stiffness on pressure is apparent.
} 
\label{fig: the cylindrical} 
\end{figure} 

At high pressures, the results of \secref{asymptotic} can be applied directly. 
For the cylindrical geometry ($\beta = 1$), the novel radius parameter becomes $\mathcal{R} = R_y\sqrt{1 + \beta} = \sqrt{2}R_y$, and the Gaussian curvature is $K = 0$. 
Substituting these forms into \eqnname{~\eqref{eqn: asymptotic novel stiffness integral 1}} and then taking the inverse of the resulting expression, we obtain the indentation stiffness of long cylindrical shells in the large-pressure limit $\p \gg 1$, 
\begin{equation} 
\ck \approx \frac{ 4\pi\sqrt{ {\kappa}Y } }{R_y}\sqrt{2}\frac{\p}{ \ln{8\p} }. 
\label{eqn: CYL. high} 
\end{equation}

\figurename{~\ref{fig: the cylindrical}} compares the numerically-evaluated stiffness for pressurized cylindrical shells (from inverting \eqnref{eqn: 1D stiffness integral} to the low-pressure (\eqnname{~\eqref{eqn: truncated CYL. low}}) and high-pressure (\eqnname{~\eqref{eqn: CYL. high}}) approximations.
We find that the analytical expressions recreate the indentation stiffness of long, thin cylindrical shells over almost all relevant pressures.
To illustrate the separation of the scales $p_\text{sc}$ and $|p_\text{c, cyl}|$, we also show the pressure using the alternate scaling $pR_y^3/\kappa$ (upper horizontal axes), using the \emph{E.~coli} cell wall parameters to connect the two scales. 
Since the resulting  F\"{o}ppl-von K\'{a}rm\'{a}n number is very large, the criterion $\ellp \ll R_y$ for the validity of our shallow-shell results is satisfied down to $\p \sim 10^{-2}$.
The inset verifies the predicted polynomial scaling of the indentation stiffness at low pressures from \eqnref{eqn: truncated CYL. low}, and shows that the zero-pressure stiffness $\ckzero$ from Ref.~\onlinecite{dePablo2003}, \eqnref{eqn:dePablo}, is approached at $\p \approx 0.005$ ($pR_y^3/\kappa \approx 10$).
Although shallow-shell theory breaks down at this low pressure, we expect that the true indentation behavior would cross over from our low-pressure expression to $\ckzero$ around this pressure value.
%%%%%%%%%%%%%%%%%%%%%%%%%%%%%%%%%%%%%%%%%%%%%%%%%%%%%%%%%%%%%%%%%%%%%%%%%%%%%%%%%%%%%%%%%%%%%%%%%%%%%%%%%%%%%%%%%%%%%%%%%%%%%%%%%%%%
\subsubsection{Stiffness Switching}

The high-pressure stiffness expression for cylinders, \eqnref{eqn: CYL. high}, is very similar to that of a sphere with the same elastic properties, radius, and internal pressure (\eqnref{eqn: SPH. asymptotic behavior}). 
To compare the relative indentation stiffness of cylinders and spheres, we compute the ratio of the two expressions: 
$$
\frac{\ck}{\sk} \approx \sqrt{2}\frac{ \ln{2\p} }{ \ln{8\p} }. 
$$
Notice that the ratio is equal to unity when $\p = 2^{2\sqrt{2} + 1} \approx 14.2$, beyond which long cylindrical shells become locally stiffer than spherical shells with the same scaled pressure.
By contrast, the sphere was stiffer at zero pressure (\eqnref{eqn: zero pressure stiffness}).
We term this phenomenon \emph{stiffness switching} between long cylindrical and spherical shells. 
In general, stiffness switching tends to occur between any pair of spheroidal shells with different asphericities upon applying an internal pressure. 
The reason is as follows.
\eqnname{~\eqref{eqn: zero pressure stiffness}} implies that at low internal pressure, the indentation stiffness is dominated by the Gaussian curvature $K = \frac{1}{R_xR_y} = \frac{1 - \beta}{ {R_y}^2 }$. 
On the other hand, $\mathcal{R}$ becomes the dominant radius parameter in the large-pressure limit, as one can rewrite the asymptotic indentation stiffness (the inverse of \eqnname{~\eqref{eqn: asymptotic scaling form}}) in a more illuminating form, 
$$
k \stackrel{\pR \gg 1}{\approx} \frac{ 4\pi\sqrt{ {\kappa}Y } }{ \mathcal{R} }\frac{\pR}{ \ln{4\pR} } 
                              = \frac{ {\pi}p\mathcal{R} }{ \ln{4\pR} } 
                              = \frac{ {\pi}pR_y\sqrt{1 + \beta} }{ \ln{4\pR} }, 
$$
whence the dominance of $\mathcal{R}$ becomes more manifest. 
Stiffness switching is hence due to the fact that $K$ and $\mathcal{R}$ have opposite $\beta$ dependences: for a fixed $R_y$, $\sqrt{K}$ is proportional to $\sqrt{1 - \beta}$, while $\mathcal{R}$ to $\sqrt{1 + \beta}$. 
The phenomenon highlights the contrasting contributions of geometry and internal pressure to the indentation stiffness of spheroidal shells.
%-----------------------------------------------------------------------------------------------------------------------------------
% \newpage 
\section{Discussion} \label{summary}

\begin{table*}[bt]
\caption{\label{tab:summary}%
Summary of analytical results for indentation stiffnesses.}
\begin{ruledtabular}
\begin{tabular}{lcc}
\textrm{Condition}&
\textrm{Parameter ranges}&
\textrm{Equation}\\
\colrule
Doubly-curved shells at zero pressure & $p = 0$, $ \sqrt{R_y t} \ll R_x \ll R_y^2/t$ & \eqref{eqn: zero pressure stiffness} or \eqref{eqn: dimensionful zero pressure stiffness} \footnote{Previously conjectured in Ref.~\onlinecite{Vella2012}.}\\
Spheroids at high internal pressure & $\pR \gg 1+ |\beta|$, $-1 < \beta \leq 1$ & \eqref{eqn: asymptotic novel stiffness integral} or \eqref{eqn: dimensionful asymptotic novel stiffness integral} \\
Spheroids under external pressure close to local instability & $1 - \frac{p}{\CP} \ll 1$, $-1 < \beta < 1$ & \eqref{eqn: critical stiffness integral} or \eqref{eqn: dimensionful critical stiffness integral} \\
Cylinders at low internal pressure & $\frac{\kappa}{R_y^3} \ll p \ll \frac{4\sqrt{\kappa Y}}{R_y^2}$, $\beta = 1$ & \eqref{eqn: truncated CYL. low} or \eqref{eqn: dimensionful truncated CYL. low} \\
Cylinders at high internal pressure & $p \gg \frac{4\sqrt{\kappa Y}}{R_y^2}$, $\beta = 1$ & \eqref{eqn: CYL. high} or \eqref{eqn: dimensionful CYL. high} \\
\end{tabular}
\end{ruledtabular}
\end{table*}

We have analyzed the linear indentation response of thin spheroidal and cylindrical shells under pressure, as a manifestation of geometric rigidity with practical applications.
While our analysis is enabled by the simplifying assumptions of shallow-shell theory, we have identified parameter regimes for which these assumptions are valid, which turn out to encompass nearly all shell geometries and pressures which allow stable prestressed states.
In addition to integral expressions for the inverse of the stiffness which can be numerically evaluated (equations~\eqref{eqn: 2D stiffness integral} and \eqref{eqn: 1D stiffness integral}), we have derived analytical expressions in various limits which rigorously validate prior results and provide easy-to-evaluate expressions for the indentation response as a function of geometric parameters, elastic properties, and pressure.
Table~\ref{tab:summary} provides a summary of these results including references to the relevant expressions.
For practical purposes, expressions written in terms of dimensionful, measurable quantities are more straightforward to utilize; in light of this, we also provide the relevant dimensionful expressions in \textbf{Appendix}~\secref{dimensionful summary}. 
We have also validated a subset of our results against data from finite-element simulations of indentation assays which were reported in Ref.~\onlinecite{Vella2012} (\figref{fig: Vella}).

Besides predictions of the indentation stiffness, our results provide insights into the nature of geometric rigidity and the influence of internal and external pressure. 
We revealed a connection between the loss of stiffness and the buckling instability of thin shells subjected to external pressure (\secref{buckling}), and showed that the behavior of the stiffness as the critical buckling pressure is approached differs qualitatively for spherical and general spheroidal shells (\secref{critical} and \figref{fig: the critical}). 
At large internal pressures, we proposed a new length scale---the distensile curvature radius, $\mathcal{R}$ (\eqnref{eqn:novel_R})---which captures the contribution of membrane prestresses to the indentation stiffness in a manner akin to how the Gaussian curvature radius $1/\sqrt{K}$ captured the geometric contribution.
The contrasting behaviors of the distensile and Gaussian curvatures as the asphericity is varied makes cylindrical shells weaker than spherical shells of the same radius at low pressures, yet stronger at high pressures---a phenomenon we termed stiffness switching.
The loss of rigidity of oblate shells with $\beta \leq -1$ at positive internal pressures is also captured by the distensile curvature taking on imaginary values.

Our result also provides new insights about the indentation stiffness of long, thin cylindrical shells.
Cylinder indentation was previously studied in the zero-pressure limit~\cite{dePablo2003} and the membrane limit which incorporated the effects of pressure-related stresses while ignoring elastic stiffness~\cite{Deng2011, Zhang2021}.
We connect these disparate regimes by exploiting a separation of pressure scales which arises for thin cylinders, which allowed us to calculate the indentation stiffness of cylindrical shells over a wide range of pressures (\secref{cylindrical shells} and \figref{fig: the cylindrical}).

Our analysis points to several promising directions for future studies.
While we focused on pressurized spheroidal shells, our approach could be used to find the indentation stiffness of any thin curved shell for which the in-plane stresses in the vicinity of the indentation point are known, as long as shallow-shell theory is applicable.
For instance, the indentation stiffness of general ellipsoids could be numerically evaluated, even away from high-symmetry points.
The approach could be extended to include the effects of a fluid or solid continuum in the shell interior, as well as material anisotropy in the shell, all of which are particularly relevant to biological structures.
It would also be interesting to analyze the indentation behavior beyond the linear regime, which would require extending the Pogorelov scaling for the energetics of large inversions of spherical shells~\cite{Pogorelov,GomezMichael2016,Baumgarten2019} to anisotropic geometries.
Understanding the large-inversion behavior would also provide insight into the post-buckling shapes of general spheroids; in this regard, the case $\beta \leq -1$ will be particularly interesting since this type of shells has two buckling states for negative and positive pressures. 
Finally, it would be interesting to extend our technique to basis functions beyond the Fourier modes we use in our analysis, which would allow us to tackle non-shallow-shells and to consider the effects of edge constraints (e.g. indentation of clamped spherical caps).
%-----------------------------------------------------------------------------------------------------------------------------------
% \newpage 
\section{Acknowledgements} 
We thank Dominic Vella for providing finite-element simulation data that was previously reported in Ref.~\onlinecite{Vella2012}, and for helpful discussions and feedback on the manuscript.
JP acknowledges support from the College of Arts and Sciences at the University of Oregon (start-up funds).
%%%%%%%%%%%%%%%%%%%%%%%%%%%%%%%%%%%%%%%%%%%%%%%%%%%%%%%%%%%%%%%%%%%%%%%%%%%%%%%%%%%%%%%%%%%%%%%%%%%%%%%%%%%%%%%
\bibliography{refs} 
%-----------------------------------------------------------------------------------------------------------------------------------
\newpage 
\appendix 
\renewcommand{\theequation}{\thesection.\arabic{equation}} 
%%%%%%%%%%%%%%%%%%%%%%%%%%%%%%%%%%%%%%%%%%%%%%%%%%%%%%%%%%%%%%%%%%%%%%%%%%%%%%%%%%%%%%%%%%%%%%%%%%%%%%%%%%%%%%%%%%%%%%%%%%%%%%%%%%%%
%%%%%%%%%%%%%%%%%%%%%%%%%%%%%%%%%%%%%%%%%%%%%%%%%%%%%%%%%%%%%%%%%%%%%%%%%%%%%%%%%%%%%%%%%%%%%%%%%%%%%%%%%%%%%%%%%%%%%%%%%%%%%%%%%%%%
%%%%%%%%%%%%%%%%%%%%%%%%%%%%%%%%%%%%%%%%%%%%%%%%%%%%%%%%%%%%%%%%%%%%%%%%%%%%%%%%%%%%%%%%%%%%%%%%%%%%%%%%%%%%%%%%%%%%%%%%%%%%%%%%%%%%
\section{A Catalogue of Dimensionful Analytical Results for Indentation Stiffnesses} \label{dimensionful summary} 
First, as a supplement to \eqnref{eqn: zero pressure stiffness}, the zero-pressure stiffness of a double-curved shell can also be written in terms of the shell material's Young's modulus (denoted by $E$), Poisson's ratio ($\upsilon$) and the shell thickness ($t$), as 
\begin{equation} 
k_{p = 0} = 8\sqrt{ \frac{\kappa Y}{R_xR_y} } 
          = \frac{4Et^2}{ \sqrt{ 3(1 - \upsilon^2) } }\frac{1}{ \sqrt{R_xR_y} }. 
\label{eqn: dimensionful zero pressure stiffness} 
\end{equation} 
When computing the indentation stiffness of a \textit{pressurized} spheroidal shell ($p \not= 0$), we believe that it is generally easier to first scale the pressure and then use the resulting nondimensionalized pressure to perform the calculations. 
Therefore, the following expressions are given in terms of the scaled pressure 
\begin{equation*} 
\p = \frac{p{R_y}^2}{ 4\sqrt{\kappa Y} } 
   = \frac{ \sqrt{ 3(1 - \upsilon^2) } }{2}\left(\frac{R_y}{t}\right)^2\frac{p}{E} 
\end{equation*} 
and $\beta = 1 - \frac{R_y}{R_x}$, the asphericity of the given shell, which is another combination of parameters that commonly shows up in the relevant indentation expressions. 

The high-pressure stiffness of spheroidal shells with $-1 < \beta \leq 1$ can be recast as 
\begin{equation} 
\ak \approx \pi pR_y\sqrt{1 + \beta} 
            \left[ 
            \ln\left(4\p\frac{1 + \beta}{ 1 + \sqrt{1 - \beta^2} }\right) 
            \right]^{-1}. 
\label{eqn: dimensionful asymptotic novel stiffness integral} 
\end{equation} 
Setting $\beta = 1$ in \eqnref{eqn: dimensionful asymptotic novel stiffness integral}, we obtain the dimensionful asymptotic stiffness of a long cylindrical shell, 
\begin{equation} 
\ack \approx \frac{\sqrt{2}\pi pR_y}{ \ln{8\p} }. 
\label{eqn: dimensionful CYL. high} 
\end{equation} 
For the long cylindrical shell, when the internal pressure is relatively low, its indentation stiffness becomes 
\begin{widetext} 
\begin{equation} 
\ck \approx 2\pi^2pR_y 
            \left( 
            11.6 \sqrt{\p} 
          - 2.66 \p^\frac{3}{2} 
          + 1.83 \p^\frac{5}{2} 
          - 0.998\p^\frac{7}{2} 
            \right)^{-1}. 
\label{eqn: dimensionful truncated CYL. low} 
\end{equation} 

Finally, for spheroidal shells with $-1 < \beta < 1$, near the critical external pressure, their stiffnesses are dictated by 
\begin{equation} 
k \approx \pi\abs{p}R_y\sqrt{ \abs{ \frac{\beta}{\p} } } 
          \frac 
          {1} 
          { 
          \displaystyle 
          \sqrt{1 - \epsilon} 
          \cdot 
          \sinh^{-1}\left( 
          \sqrt{ \frac{2}{ \frac{4}{\pi^2}\frac{1}{ \abs{\beta} } + 1 + f(\beta) } }\frac{1}{ \sqrt{\epsilon} } 
          \right) 
          }, 
          \quad 
          (p, \p < 0) 
\label{eqn: dimensionful critical stiffness integral} 
\end{equation} 
where 
\begin{equation*} 
\epsilon = \begin{cases} 
           1 - \frac{1 + \beta}{1 - \beta}\abs{\p}, & \text{for}\ 0 \leq \beta < 1, \\[0.25 em] 
           1 - \abs{\p},                            & \text{for}\ -1 < \beta < 0, 
           \end{cases} 
           \quad \text{and} \quad 
f(\beta) = \begin{cases} 
           \left(1 - \frac{4}{\pi^2}\right)\beta,   & \text{for}\ 0 \leq \beta < 1, \\[0.25 em] 
           0,                                       & \text{for}\ -1 < \beta < 0. 
           \end{cases} 
\end{equation*} 
\end{widetext} 
%-----------------------------------------------------------------------------------------------------------------------------------
\end{document}